\documentclass[aps,prd,twocolumn,superscriptaddress,showpacs]{revtex4}
\usepackage[dvips]{graphicx}
\usepackage{amsmath,amssymb,times}
\usepackage{braket}

\newcommand{\bequ}{\begin{equation}}
\newcommand{\eequ}{\end{equation}}
\newcommand{\bea}{\begin{eqnarray}}
\newcommand{\eea}{\end{eqnarray}}

\newcommand{\la}{\left\langle}

\def\la{\mathrel{\mathpalette\fun <}}
\def\ga{\mathrel{\mathpalette\fun >}}
\def\fun#1#2{\lower3.6pt\vbox{\baselineskip0pt\lineskip.9pt
  \ialign{$\mathsurround=0pt#1\hfil##\hfil$\crcr#2\crcr\sim\crcr}}}

\DeclareSymbolFont{boldletters}{OML}{cmm} {b}{it}
\DeclareSymbolFontAlphabet{\mathbit}{boldletters}
\DeclareMathSymbol{\alpha}{\mathalpha}{letters}{"0B}
\DeclareMathSymbol{\beta}{\mathalpha}{letters}{"0C}
\DeclareMathSymbol{\gamma}{\mathalpha}{letters}{"0D}
\DeclareMathSymbol{\delta}{\mathalpha}{letters}{"0E}
\DeclareMathSymbol{\epsilon}{\mathalpha}{letters}{"0F}
\DeclareMathSymbol{\zeta}{\mathalpha}{letters}{"10}
\DeclareMathSymbol{\eta}{\mathalpha}{letters}{"11}
\DeclareMathSymbol{\theta}{\mathalpha}{letters}{"12}
\DeclareMathSymbol{\iota}{\mathalpha}{letters}{"13}
\DeclareMathSymbol{\kappa}{\mathalpha}{letters}{"14}
\DeclareMathSymbol{\lambda}{\mathalpha}{letters}{"15}
\DeclareMathSymbol{\mu}{\mathalpha}{letters}{"16}
\DeclareMathSymbol{\nu}{\mathalpha}{letters}{"17}
\DeclareMathSymbol{\xi}{\mathalpha}{letters}{"18}
\DeclareMathSymbol{\pi}{\mathalpha}{letters}{"19}
\DeclareMathSymbol{\rho}{\mathalpha}{letters}{"1A}
\DeclareMathSymbol{\sigma}{\mathalpha}{letters}{"1B}
\DeclareMathSymbol{\tau}{\mathalpha}{letters}{"1C}
\DeclareMathSymbol{\upsilon}{\mathalpha}{letters}{"1D}
\DeclareMathSymbol{\phi}{\mathalpha}{letters}{"1E}
\DeclareMathSymbol{\chi}{\mathalpha}{letters}{"1F}
\DeclareMathSymbol{\psi}{\mathalpha}{letters}{"20}
\DeclareMathSymbol{\omega}{\mathalpha}{letters}{"21}
\DeclareMathSymbol{\varepsilon}{\mathalpha}{letters}{"22}
\DeclareMathSymbol{\vartheta}{\mathalpha}{letters}{"23}
\DeclareMathSymbol{\varpi}{\mathalpha}{letters}{"24}
\DeclareMathSymbol{\varrho}{\mathalpha}{letters}{"25}
\DeclareMathSymbol{\varsigma}{\mathalpha}{letters}{"26}
\DeclareMathSymbol{\varphi}{\mathalpha}{letters}{"27}
\DeclareMathSymbol{\Gamma}{\mathalpha}{letters}{"00}
\DeclareMathSymbol{\Delta}{\mathalpha}{letters}{"01}
\DeclareMathSymbol{\Theta}{\mathalpha}{letters}{"02}
\DeclareMathSymbol{\Lambda}{\mathalpha}{letters}{"03}
\DeclareMathSymbol{\Xi}{\mathalpha}{letters}{"04}
\DeclareMathSymbol{\Pi}{\mathalpha}{letters}{"05}
\DeclareMathSymbol{\Sigma}{\mathalpha}{letters}{"06}
\DeclareMathSymbol{\Upsilon}{\mathalpha}{letters}{"07}
\DeclareMathSymbol{\Phi}{\mathalpha}{letters}{"08}
\DeclareMathSymbol{\Psi}{\mathalpha}{letters}{"09}
\DeclareMathSymbol{\Omega}{\mathalpha}{letters}{"0A}

\begin{document}
\preprint{SAGA-HE-255-09}
\title{QCD phase diagram at finite baryon and isospin chemical potentials}

\author{Takahiro Sasaki}
\email[]{sasaki@phys.kyushu-u.ac.jp}
\affiliation{Department of Physics, Graduate School of Sciences, Kyushu University,
             Fukuoka 812-8581, Japan}

\author{Yuji Sakai}
\email[]{sakai@phys.kyushu-u.ac.jp}
\affiliation{Department of Physics, Graduate School of Sciences, Kyushu University,
             Fukuoka 812-8581, Japan}

\author{Hiroaki Kouno}
\email[]{kounoh@cc.saga-u.ac.jp}
\affiliation{Department of Physics, Saga University,
             Saga 840-8502, Japan}

\author{Masanobu Yahiro}
\email[]{yahiro@phys.kyushu-u.ac.jp}
\affiliation{Department of Physics, Graduate School of Sciences, Kyushu University,
             Fukuoka 812-8581, Japan}

\date{\today}

\begin{abstract}
The phase structure of two-flavor QCD 
is explored for thermal systems with finite 
baryon- and isospin-chemical potentials, 
$\mu_{\rm B}$ and $\mu_{\rm iso}$, by 
using the Polyakov-loop extended Nambu--Jona-Lasinio (PNJL) model. 
The PNJL model with the scalar-type eight-quark interaction can reproduce 
lattice QCD data 
at not only $\mu_{\rm iso}=\mu_{\rm B}=0$ 
but also $\mu_{\rm iso}>0$ and $\mu_{\rm B}=0$. 
In the $\mu_{\rm iso}$-$\mu_{\rm B}$-$T$ space, 
where $T$ is temperature, 
the critical endpoint of the chiral phase transition in the 
$\mu_{\rm B}$-$T$ plane at $\mu_{\rm iso}$=0 moves to 
the tricritical point of the pion-superfluidity phase transition 
in the $\mu_{\rm iso}$-$T$ plane at $\mu_{\rm B}$=0 
as $\mu_{\rm iso}$ increases. 
The thermodynamics at small $T$ is controlled by 
$\sqrt{\sigma^2+\pi^2}$ defined by the chiral and pion condensates, 
$\sigma$ and $\pi$. 
\end{abstract}

\maketitle

\section{Introduction}
\label{Introduction}

The phase diagram of quantum chromodynamics (QCD) is 
the key to understanding not only natural 
phenomena such as compact stars and the early Universe but also 
laboratory experiments such as relativistic heavy-ion collisions. 
Quantitative calculations of the phase diagram from 
the first-principle lattice QCD (LQCD) have the well-known sign problem 
when the baryon chemical potential ($\mu_{\rm B}$) is real~\cite{Kogut}; 
here, $\mu_{\rm B}$ is related to the quark-number chemical potential 
$\mu_{\rm q}$ as $\mu_{\rm B}=3\mu_{\rm q}$. 
Several approaches have been proposed so far 
to circumvent the difficulty; 
for example, the reweighting method~\cite{Fodor}, 
the Taylor expansion method~\cite{Allton} and 
the analytic continuation from imaginary $\mu_{\rm q}$ 
to real $\mu_{\rm q}$~\cite{FP,Elia,Chen}. 
However, those are still far from perfection particularly at 
$\mu_{\rm q}/T \ga 1$, where $T$ is temperature. 

As an approach complementary to LQCD, 
we can consider effective models such as the  
Nambu--Jona-Lasinio (NJL) model~
\cite{NJ1,AY,KKKN,Fujii,Kashiwa1,He} and 
the Polyakov-loop extended Nambu--Jona-Lasinio (PNJL) 
model~\cite{Meisinger,Fukushima,Fukushima2,Ghos,Megias,Ratti,Ciminale,
Rossner,Hansen,Sasaki,Schaefer,Zhang,Mukherjee,Kashiwa2,Fu,Abuki,AF,Hell,
Sakai1,Sakai2,Kashiwa5}. 
The NJL model describes the chiral symmetry breaking, but not 
the confinement mechanism. 
The PNJL model is extended so as 
to treat both the mechanisms ~\cite{Fukushima} approximately by 
considering the Polyakov-loop in addition to the chiral condensate 
as ingredients of the model.

In the NJL-type models, 
the input parameters are usually determined from the 
pion mass and the pion decay constant 
at vacuum ($\mu_{\rm q}=0$ and $T = 0$). 
Some of the models have the scalar-type eight-quark interaction. 
The strength of the 
interaction is adjusted to LQCD data at finite $T$~\cite{Sakai1}, since 
the sigma-meson mass at vacuum related to the interaction 
has a large error bar and then ambiguous~\cite{Schaefer2}. 
It is then highly nontrivial whether the models predict properly dynamics 
of QCD at finite $\mu_{\rm q}$.  
This should be tested from QCD. 
Fortunately, this is possible at imaginary $\mu_{\rm q}$, 
since LQCD has no sign problem there.
In Ref.~\cite{Sakai2}, it was shown that the PNJL model can reproduce 
LQCD data at imaginary $\mu_{\rm q}$. 
QCD has the Roberge-Weiss periodicity and 
the Roberge-Weiss transition~\cite{RW} in the imaginary $\mu_{\rm q}$ region, 
because of the extended $\mathbb{Z}_3$ symmetry~\cite{Sakai1,Sakai2,Kashiwa5}.  The PNJL model can reproduce these, since it 
has the symmetry~\cite{Sakai1}. 
In the real $\mu_{\rm q}$ region, as an important result, a phase diagram 
is predicted by the PNJL model 
with the parameter set~\cite{Sakai2} determined from 
the LQCD data at imaginary $\mu_{\rm q}$. 
The PNJL prediction shows that in the $\mu_{\rm q}$-$T$ plane at 
$\mu_{\rm I}$=0 
there appears a critical endpoint (CEP), that is, a second-order 
critical point where a first-order chiral phase-transition line terminates.

A similar test of the PNJL model is possible for 
finite isospin-chemical potential ($\mu_{\rm iso}$)~\cite{Son-Stephanov}, since LQCD has no sign problem there; 
for later convenience, we use the ``modified" isospin-chemical 
potential $\mu_{\rm I}=\mu_{\rm iso}/2$ instead of $\mu_{\rm iso}$. 
LQCD data are available 
for both real~\cite{Kogut2} and imaginary~\cite{Cea,D'Elia} 
$\mu_{\rm iso}$. 
The PNJL model has already been applied to the real 
$\mu_{\rm I}$~\cite{Zhang,Mukherjee} and the imaginary 
$\mu_{\rm I}$ case~\cite{Sakai3} with 
success in reproducing the LQCD data. 
The PNJL calculation at real $\mu_{\rm I}$ shows that 
in the $\mu_{\rm I}$-$T$ plane at $\mu_{\rm q}=0$ 
there exists a first-order pion-superfluidity phase-transition line 
connected to a second-order 
pion-superfluidity phase-transition line; 
the connecting point is a tricritical point (TCP) by definition. 
Real $\mu _{\rm I}$ dependence of QCD phase diagram is investigated
 in other models such as chiral perturbation 
 theory~\cite{Son-Stephanov}, the strong coupling QCD \cite{Nishida} 
 and so on \cite{Fraga}.

The CEP in the $\mu_{\rm q}$-$T$ plane at $\mu_{\rm I}=0$ is 
important as 
a good indicator of the chiral and deconfinement 
phase transitions in collider experiments at 
Helmholtzzentrum f\"{u}r Schwerionenforschung GmbH (GSI), Super Proton Synchrotron (SPS)~\cite{Anticic,Anticic2}, Relativistic Heavy Ion Collider (RHIC)~\cite{Lacey,Sorensen} 
and LHC~\cite{Nayak}.
In the measurements, $\mu_{\rm I}$ is not zero generally. 
It is then interesting to see how critical points 
such as CEP and TCP are located in the $\mu_{\rm q}$-$\mu_{\rm I}$-$T$ space.  

In this paper, we draw the phase diagram of two-flavor QCD
in the $\mu_{\rm I}$-$\mu_{\rm q}$-$T$ space by using the PNJL model. 
Following our previous paper~\cite{Sakai2}, 
we introduce the scalar-type eight-quark interaction to 
reproduce LQCD data on 
thermal systems at $\mu_{\rm q}=\mu_{\rm I}=0$. 
The scalar-type eight-quark interaction is a next-to-leading order correction 
in the power counting rule based on mass dimension. 
First, we will show that the PNJL model 
with the parameter set thus determined also reproduces 
LQCD data on thermal systems at $\mu_{\rm I} >0 $ and $\mu_{\rm q}=0$. 
After confirming the reliability of the present PNJL model, 
we will predict locations of CEP and TCP 
in the $\mu_{\rm I}$-$\mu_{\rm q}$-$T$ space.

In Sec.~\ref{PNJL}, the PNJL model is recapitulated. 
In Sec.~\ref{Numerical-results}, 
the PNJL calculation is compared with LQCD data for thermal systems 
at $\mu_{\rm I} >0 $ and $\mu_{\rm q}=0$, and 
the phase diagram is explored 
in the $\mu_{\rm I}$-$\mu_{\rm q}$-$T$ space. 
Properties of the susceptibilities near CEP and TCP are analyzed. 
Section~\ref{Summary} is devoted to a summary.

\section{PNJL model}
\label{PNJL}
The two-flavor PNJL Lagrangian in Euclidean space-time is 
\bea
\mathcal{L}&=&
\bar{q}(\gamma_\nu D^\nu - \gamma_4{\hat \mu} + {\hat m}_0 )q 
+G_{\rm s}\left[
(\bar{q}q)^{2}+(\bar{q}i\gamma_{5} \vec{\tau} q)^{2}
\right]
 \notag \\
&+&
G_{8}\left[
(\bar{q}q)^{2}+(\bar{q}i\gamma_{5} \vec{\tau} q)^{2}
\right] ^{2}
- {\cal U}(\Phi [A],{\Phi} [A]^*,T), 
\label{eq:E1}
\eea
where $D^\nu=\partial^\nu+iA^\nu$ and 
$A^\nu=\delta^{\nu}_{0}gA^0_a{\lambda^a\over{2}}$
with the gauge field $A^\nu_a$, 
the Gell-Mann matrix $\lambda_a$ and the gauge coupling $g$. 
In the NJL sector, 
$G_{\rm s}$ denotes the coupling constant of the scalar-type 
four-quark interaction and 
$G_{8}$ stands for that of the scalar-type 
eight-quark interaction~\cite{Sakai2,Bhattacharyya,Osipov}.
The Polyakov-potential ${\cal U}$, defined in (\ref{eq:E13}), 
is a function of the Polyakov-loop $\Phi$ and its Hermitian 
conjugate $\Phi^*$. 

The chemical potential matrix ${\hat \mu}$ is 
defined by ${\hat \mu}={\rm diag}(\mu_u, \mu_d)$ with 
the $u$-quark ($d$-quark) number chemical potential $\mu_{u}$ ($\mu_{d}$), 
while ${\hat m}_0={\rm diag}(m_0, m_0)$. 
This is 
equivalent to introducing the baryon and isospin-chemical potentials, 
$\mu_{\rm B}$ and $\mu_{\rm iso}$, 
coupled, respectively, to the baryon charge ${\bar B}$ and to the isospin 
charge ${\bar I_3}$:
\bea
{\hat \mu}=\mu_{\rm q} \tau_0 + \mu_{\rm I} \tau_3 
\eea
with 
\begin{align}
\mu_{\rm q}=\frac{\mu_{u}+\mu_{d}}{2}=\frac{\mu_{\rm B}}{3},
~~\mu_{\rm I}=\frac{\mu_{u}-\mu_{d}}{2}=\frac{\mu_{\rm iso}}{2} , 
\end{align}
where $\tau_0$ is the unit matrix and $\tau_i$ ($i=1, 2, 3$) 
are the Pauli matrices in flavor space. 
Note that $\mu_{\rm q}$ is the quark chemical potential and 
$\mu_{\rm I}$ is half the isospin-chemical potential ($\mu_{\rm iso}$). 
In the limit of $m_0=\mu_{\rm I}=0$, 
the PNJL Lagrangian has the $SU_{\rm L}(2) \times SU_{\rm R}(2)
\times U_{\rm v}(1) \times SU_{\rm c}(3)$  symmetry. 
For $m_0 \neq 0$ and $\mu_{\rm I} \neq 0$, 
it is reduced to $U_{\rm I_3}(1) \times U_{\rm v}(1) \times SU_{\rm c}(3)$.

The Polyakov-loop operator $\hat{\Phi}$ and its Hermitian conjugate 
$\hat{\Phi}^{\dagger}$ are defined as
\begin{eqnarray}
\hat{\Phi}        &=& {1\over{N}} {\rm Tr} L ,~~~~
\hat{\Phi}^{\dagger}  = {1\over{N}} {\rm Tr}L^\dag ,
\end{eqnarray}
with
\begin{eqnarray}
L({\bf x})  &=& {\cal P} \exp\Bigl[
                {i\int^\beta_0 d \tau A_4({\bf x},\tau)}\Bigr],
\end{eqnarray}
where ${\cal P}$ is the path ordering and $A_4 = i A^0 $. 
In the PNJL model, the vacuum expectation values,  
$\Phi=\langle \hat{\Phi} \rangle$ and 
$\Phi^{*}=\langle \hat{\Phi}^{\dagger}  \rangle$, 
are treated as classical variables.  
In the Polyakov gauge, $L$ can be written in a diagonal form 
in color space~\cite{Fukushima}: 
\begin{align}
L 
=  e^{i \beta (\phi_3 \lambda_3 + \phi_8 \lambda_8)}
= {\rm diag} (e^{i \beta \phi_a},e^{i \beta \phi_b},
e^{i \beta \phi_c} ),
\label{eq:E6}
\end{align}
where $\phi_a=\phi_3+\phi_8/\sqrt{3}$, $\phi_b=-\phi_3+\phi_8/\sqrt{3}$
and $\phi_c=-(\phi_a+\phi_b)=-2\phi_8/\sqrt{3}$. 

The Polyakov-loop $\Phi$ is an exact order parameter of the spontaneous 
${\mathbb Z}_3$ symmetry breaking in the pure gauge theory.
Although the ${\mathbb Z}_3$ symmetry is not exact 
in the system with dynamical quarks, it still seems to be a good indicator of 
the deconfinement phase transition. 
Therefore, we use $\Phi$ to define the deconfinement phase transition.

The spontaneous breakings of the chiral and the $U_{\rm I_3}(1)$ symmetry are 
described 
by the chiral condensate $\sigma = \langle \bar{q}q \rangle$ and the charged 
pion condensate~\cite{Zhang}
\begin{align}
\pi^{\pm}=\frac{\pi}{\sqrt{2}}e^{\pm i \varphi}
=\langle \bar{q}i \gamma_5 \tau_{\pm}q \rangle ,
\label{charged-pion}
\end{align}
where $\tau_{\pm}=(\tau_1\pm i \tau_2)/\sqrt{2}$.
Since the phase $\varphi$ represents the direction 
of the $U_{\rm I_3}(1)$ symmetry breaking, 
we take $\varphi=0$ for convenience. The pion 
condensate is then expressed by 
\begin{align}
\pi=\langle \bar{q}i \gamma_5 \tau_{1}q \rangle.
\label{pion}
\end{align}
Making the mean field (MF) approximation~\cite{Kashiwa1,Zhang}, 
one can obtain the MF Lagrangian as 
\begin{align}
 {\cal L}_{\rm MF}  &=& {\bar q}(\gamma_\nu D^\nu - \gamma_4{\hat \mu} + 
              M\tau_0 + N i \gamma_5 \tau_{1})q  ~~~~~~\notag\\
            &\hspace{3mm}&  - G_{\rm s}[\sigma^2 +\pi^2] 
              - 3G_{8}(\sigma ^{2} + \pi^{2})^{2}
              - {\cal U} \quad
             \label{MF-L}
\end{align}
with 
\begin{eqnarray}
M&=&m_{0} - 2G_{\rm s} \sigma -4G_{8}\sigma (\sigma ^{2}+\pi ^{2}) , 
\label{MM} \\
N&=&- 2G_{\rm s} \pi -4G_{8}\pi (\sigma ^{2}+\pi ^{2}) . 
\label{NN}
\end{eqnarray}
Performing the path integral in the PNJL partition function 
\begin{align}
Z_{\rm PNJL}=\int Dq D\bar{q}
\exp\left[ - \int d^4 x {\cal L}_{\rm MF} \right] , 
\label{PNJL-Z}
\end{align}
we can get the thermodynamic potential $\Omega$ 
(per unit volume), 
\begin{align}
\Omega &=-T\ln(Z_{\rm PNJL})/V
= -2\sum_{i=\pm}\int \frac{d^3{\rm p}}{(2\pi)^3}
         \Bigl[ 3 E_{i}({\rm p}) \nonumber\\
       & + \frac{1}{\beta}
         \ln~ [1 + 3(\Phi+\Phi^{*} e^{-\beta E_{i}^-({\bf p})}) 
        e^{-\beta E_{i}^-({\bf p})}+ e^{-3\beta E_{i}^- ({\bf p})}]
         \nonumber\\
       & + \frac{1}{\beta} 
           \ln~ [1 + 3(\Phi^{*}+{\Phi e^{-\beta E_{i}^+({\bf p})}}) 
            e^{-\beta E_{i}^+({\bf p})}+ e^{-3\beta E_{i}^+({\bf p})}]
           \Bigl]\nonumber\\
       & +G_{\rm s}[\sigma^2 +\pi^2]
       + 3G_{8}(\sigma ^{2} + \pi^{2})^{2}
       +{\cal U} 
\label{eq:E12-pi} 
\end{align}
with $E_{\pm}^\pm({\rm p})=E_{\pm}({\rm p})\pm \mu_{\rm q}$, 
where  
\bea
E_{\pm}({\rm p})
=\sqrt{(E({\rm p})\pm\mu_{\rm I})^2+N^2} 
\eea
for $E({\rm p})=\sqrt{{\bf p}^2+M^2}$. 
On the right-hand side of \eqref{eq:E12-pi}, only the first term diverges, and 
it is then regularized by the three-dimensional momentum 
cutoff $\Lambda$~\cite{Fukushima,Ratti}. 

We use ${\cal U}$ of Ref.~\cite{Rossner} that is fitted to LQCD data 
in the pure gauge theory at finite $T$~\cite{Boyd,Kaczmarek}: 
\begin{align}
&{\cal U} = T^4 \Bigl[-\frac{a(T)}{2} {\Phi}^*\Phi\notag\\
      &~~~~~+ b(T)\ln(1 - 6{\Phi\Phi^*}  + 4(\Phi^3+{\Phi^*}^3)
            - 3(\Phi\Phi^*)^2 )\Bigr], \label{eq:E13}\\
&a(T)   = a_0 + a_1\Bigl(\frac{T_0}{T}\Bigr)
                 + a_2\Bigl(\frac{T_0}{T}\Bigr)^2,
 ~~~b(T)=b_3\Bigl(\frac{T_0}{T}\Bigr)^3 , \label{eq:E14}
\end{align}
where parameters are summarized in Table \ref{table-para}.  
The Polyakov potential yields a first-order deconfinement phase transition at 
$T=T_0$ in the pure gauge theory.
The original value of $T_0$ is $270$ MeV determined from the pure gauge 
LQCD data, but the PNJL model with this value of $T_0$ yields a larger 
value of the pseudocritical temperature $T_\mathrm{c}$ 
at zero chemical potential than $T_{\rm c}=173 \pm 8$~MeV that 
the full LQCD simulation~\cite{Karsch3,Karsch4,Kaczmarek2} predicts. 
Therefore, we reset $T_0$ to 212~MeV~\cite{Sakai2} so as to 
reproduce the LQCD result. 

\begin{table}[h]
\begin{center}
\begin{tabular}{llllll}
\hline \hline
~~~~~$a_0$~~~~~&~~~~~$a_1$~~~~~&~~~~~$a_2$~~~~~&~~~~~$b_3$~~~~~
\\
\hline
~~~~3.51 &~~~~-2.47 &~~~~15.2 &~~~~-1.75\\
\hline \hline
\end{tabular}
\caption{
Summary of the parameter set in the Polyakov-potential sector 
determined in Ref.~\cite{Rossner}. 
All parameters are dimensionless. 
}
\label{table-para}
\end{center}
\end{table}

The classical variables $X=\Phi$, ${\Phi}^*$, $\sigma$, and $\pi$ 
are determined by the stationary conditions 
\begin{align}
\partial \Omega/\partial X=0. 
\label{eq:SC}
\end{align}
The solutions to the stationary conditions do not give 
the global minimum of $\Omega$ 
necessarily. There is a possibility 
that they yield a local minimum or even 
a maximum. We then have checked that the solutions yield 
the global minimum when  
the solutions 
$X(T,\mu_{\rm q},\mu_{\rm I})$
are inserted into (\ref{eq:E12-pi}).

In this work, first-order transitions are defined by 
(approximate) order parameters, $\sigma$, $\pi$ and $\Phi$ in their 
discontinuities. When the susceptibility of one of the order parameters 
diverges, 
we regard it as a second-order transition of the order parameter. 
For crossover, the pseudocritical point is determined by a peak of 
the susceptibility. When the susceptibility has two peaks and 
it is not clear which peak should be taken, we do not plot a phase boundary 
to avoid the confusion.

Table \ref{NJL-para} shows parameters in the NJL sector used in 
the present analyses. 
As shown in Ref.~\cite{Sakai2}, set A can reproduce 
not only the pion decay constant $f_{\pi}=93.3$~MeV and the pion mass 
$M_{\pi}=138$~MeV at vacuum ($T=\mu_{\rm q}=\mu_{\rm I}=0$)
but also $T_{\rm c}=173 \pm 8$~MeV~\cite{Karsch4,Karsch3,Kaczmarek2} 
at finite temperature ($T > 0$ and $\mu_{\rm q}=\mu_{\rm I}=0$). 
For this reason, we take this parameter set in this paper. 
For comparison, we also use set B with no scalar-type eight-quark interaction. 
This parameter set also reproduces the pion mass and the pion decay constant 
correctly, but not LQCD data at finite temperature 
($T > 0$ and $\mu_{\rm q}=\mu_{\rm I}=0$). 
The sigma-meson mass $M_{\sigma}$ is 526 (680)~[MeV] for set A (B).

\begin{table}[h]
\begin{center}
\begin{tabular}{lcccc}
\hline \hline
Set~&$G_s$ & $G_8$ & $m_0$ & $\Lambda$
\\
\hline
A&~$4.673~[{\rm GeV}^{-2}]$~& ~$452.12~[{\rm GeV}^{-8}]$~ & ~$5.5~[{\rm MeV}]$~ & ~$631.5~[{\rm MeV}]$~ \\
B&~$5.498~[{\rm GeV}^{-2}]$~& ~$0$~ & ~$5.5~[{\rm MeV}]$~ & ~$631.5~[{\rm MeV}]$~ \\
\hline \hline
\end{tabular}
\caption{
Summary of parameters in the NJL sector. 
Here, $T_0=212$~MeV for both the sets. 
}
\label{NJL-para}
\end{center}
\end{table}

\section{Numerical results}
\label{Numerical-results}

The phase structure in the $\mu_{\rm I}$-$\mu_{\rm q}$-$T$ space 
is explored by the PNJL model with the eight-quark interaction. 

\subsection{Phase structure in the $\mu_{\rm I}$-$T$ plane at $\mu_{\rm q}=0$}
\label{the IT plane}

LQCD data are available in the $\mu_{\rm I}$-$T$ plane 
at $\mu_{\rm q}=0$~\cite{Kogut2}, 
since LQCD has no sign problem there. In QCD, it is known~\cite{Son-Stephanov} 
that at zero $T$, a second-order phase transition occurs 
at $\mu_{\rm I}=M_{\pi}/2$ from the normal ($\pi=0$) 
to the pion-superfluidity phase ($\pi \neq 0$); 
this will be understood in subsection \ref{the IQ plane} also by using 
the PNJL model with the eight-quark interaction. 
The critical chemical potential $\mu_{\rm c}$ of 
the pion-superfluidity phase transition is 
$\mu_{\rm c}=0.57/a$ in LQCD calculation with a lattice spacing $a$, 
while it is $\mu_{\rm c}=M_{\pi}/2=69$~[MeV] in the PNJL calculation. 
In the LQCD data, $\mu_{\rm I}$ is then normalized as 
$\mu_{\rm c}=69$~[MeV]. This makes it possible 
to compare the PNJL calculation with the LQCD data.

\begin{figure}[htbp]
\begin{center}
 \includegraphics[width=0.5\textwidth]{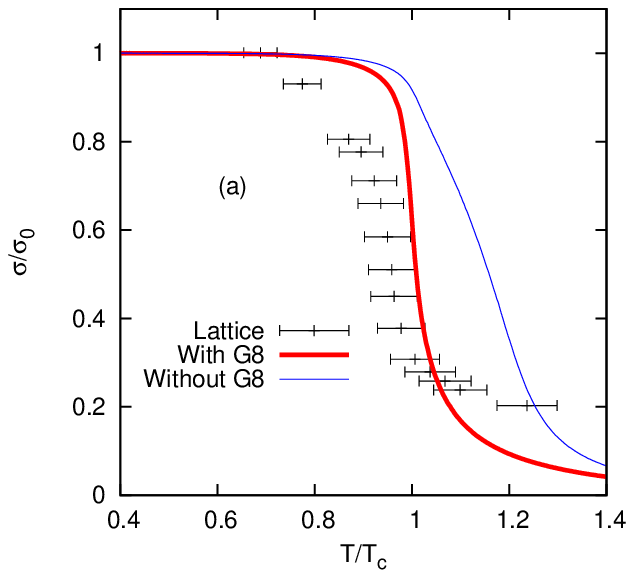}
 \includegraphics[width=0.5\textwidth]{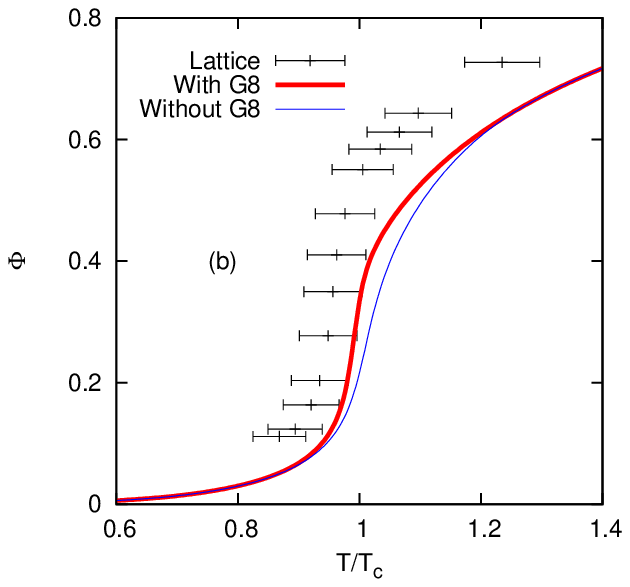} 
\end{center}
\caption{(color online). $T$ dependence of 
(a) chiral condensate $\sigma$ and (b) Polyakov-loop $\Phi$ 
 at $\mu_{\rm I}=0.96\mu_{\rm c}=66$~[MeV] and $\mu_{\rm q}=0$. 
 Here, $\sigma$ is normalized by the value $\sigma_0$ at vacuum and 
 $T$ is also normalized by $T_c=173[{\rm MeV}]$.  
 The thick (thin) solid curves represent the PNJL results with (without) 
 the scalar-type eight-quark interaction; 
 Lattice data ($+$) are taken from Ref.~\cite{Kogut2}. The lattice data 
 are plotted with 10 \% error bar, since lattice calculations have 
 10 \% error in determining $T_\mathrm{c}$~\cite{Karsch4}. 
  }
\label{T-dep-muq0}
\end{figure}

First, we consider the normal phase by taking a case of 
$\mu_{\rm I}=0.96\mu_{\rm c}=66$~[MeV]. 
Figure~\ref{T-dep-muq0} presents $\sigma$ and $\Phi$ 
as a function of $T/T_\mathrm{c}$, 
where $\sigma$ is normalized by the value $\sigma_0$ at zero $T$. 
LQCD data are plotted by plus ($+$) symbols with 10~\% error bar; 
LQCD data of Refs.~\cite{Kogut2} have only 
small errors, but we have added 10 \% error 
that comes from LQCD data~\cite{Karsch4} on 
the pseudocritical temperature $T_\mathrm{c}$ at zero quark and isospin-chemical potentials. 
The PNJL result with 
the scalar-type eight-quark interaction (the thick-solid curve) 
is consistent with the LQCD data; 
note that the present model has no free parameter. 
If the scalar-type eight-quark interaction is switched off from 
the PNJL model, the result (the thin-solid curve) 
deviates sizably from the LQCD data 
particularly on $\sigma$. This indicates that 
the scalar-type eight-quark interaction is inevitable. 

We use the dimensionless susceptibility matrix~\cite{Fukushima2,Kashiwa2} 
\bea
\tilde{\chi} &=&C^{-1}
\label{chi-4}
\eea
defined by the dimensionless curvature matrix
\begin{eqnarray}
C&=&
\left(
\begin{array}{cccc}
c_{\pi\pi}&
c_{\pi\sigma}&
c_{\pi\Phi}&
c_{\pi\bar{\Phi}}
\\
c_{\sigma\pi}&
c_{\sigma\sigma}&
c_{\sigma\Phi}&
c_{\sigma\bar{\Phi}}
\\
c_{\Phi\pi}&
c_{\Phi\sigma}&
c_{\Phi\Phi}&
c_{\Phi\bar{\Phi}}
\\
c_{\bar{\Phi}\pi}&
c_{\bar{\Phi}\sigma}&
c_{\bar{\Phi}\Phi}&
c_{\bar{\Phi}\bar{\Phi}}
\end{array}
\right)
\nonumber \\
&=&
\left(
\begin{array}{cccc}
{}^{~}T^{2}\Omega_{\pi\pi}&
{}^{~}T^{2}\Omega_{\pi\sigma}&
T^{-1}\Omega_{\pi\Phi}&
T^{-1}\Omega_{\pi\bar{\Phi}}
\\
{}^{~}T^{2}\Omega_{\sigma\pi}&
{}^{~}T^{2}\Omega_{\sigma\sigma}&
T^{-1}\Omega_{\sigma\Phi}&
T^{-1}\Omega_{\sigma\bar{\Phi}}
\\
T^{-1}\Omega_{\Phi\pi}&
T^{-1}\Omega_{\Phi\sigma}&
T^{-4}\Omega_{\Phi\Phi}&
T^{-4}\Omega_{\Phi\bar{\Phi}}
\\
T^{-1}\Omega_{\bar{\Phi}\pi}&
T^{-1}\Omega_{\bar{\Phi}\sigma}&
T^{-4}\Omega_{\bar{\Phi}\Phi}&
T^{-4}\Omega_{\bar{\Phi}\bar{\Phi}}
\end{array}
\right),~~~~
\label{curvature}
\end{eqnarray}
where $\Omega_{xy}=\partial^{2}\Omega/\partial x\partial y$ for 
$x, y=\sigma,\pi,\Phi,\bar{\Phi}$. 
The susceptibilities thus defined are dimensionless.  
For simplicity, we take the following shorthand notation: 
$\tilde{\chi} _{\sigma}=\tilde{\chi}_{\sigma\sigma},
\tilde{\chi} _{\pi}=\tilde{\chi}_{\pi\pi},\tilde{\chi} _{\Phi}=\tilde{\chi}_{\Phi\Phi}$.

\begin{figure}[htbp]
\begin{center}
 \includegraphics[width=0.5\textwidth]{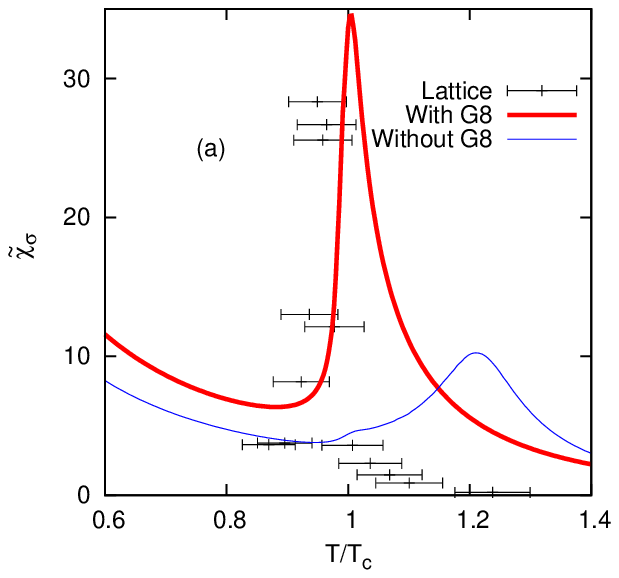}
 \includegraphics[width=0.5\textwidth]{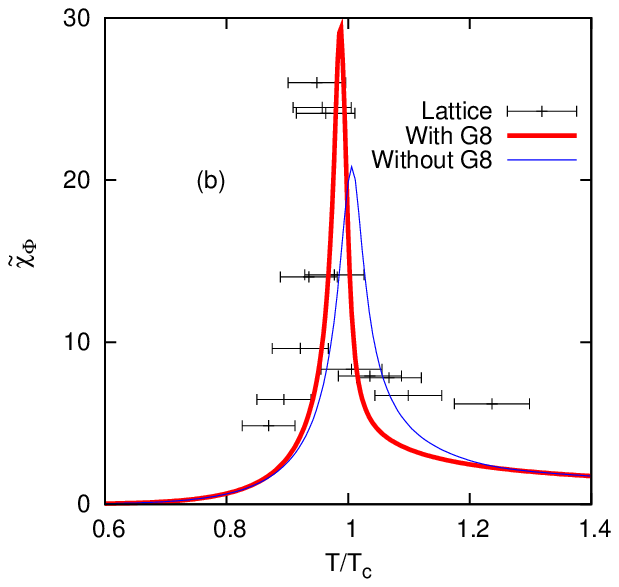}
\end{center}
\caption{(color online). $T$ dependence of 
(a) chiral and (b) Polyakov-loop susceptibility 
at $\mu_{\rm I}=0.96\mu_{\rm c}=66$~[MeV] and $\mu_{\rm q}=0$. 
See Fig.~\ref{T-dep-muq0} for the definition of lines and LQCD data. 
Since the susceptibilities of LQCD are obtained in arbitrary units, 
the magnitudes are then rescaled to fit the corresponding 
thick-solid curves, respectively. 
  }
\label{T-dep-muq0-2}
\end{figure}

Figure~\ref{T-dep-muq0-2} presents the chiral and the Polyakov-loop 
susceptibility, $\tilde{\chi}_{\sigma}$ and $\tilde{\chi}_{\Phi}$, 
as a function of $T$. 
The PNJL model with the scalar-type eight-quark interaction 
(the thick-solid curve) gives a better agreement with the LQCD data than 
the PNJL model without the scalar-type eight-quark interaction 
(the thin-solid curve). 
The present analysis for finite $\mu_{\rm I}$ is parameter free. 
Therefore, 
the reasonable agreement between the PNJL model 
with the eight-quark interaction and the LQCD data indicates that 
the PNJL model with the eight-quark interaction is reliable.

\begin{figure}[htbp]
\begin{center}
 \includegraphics[width=0.5\textwidth]{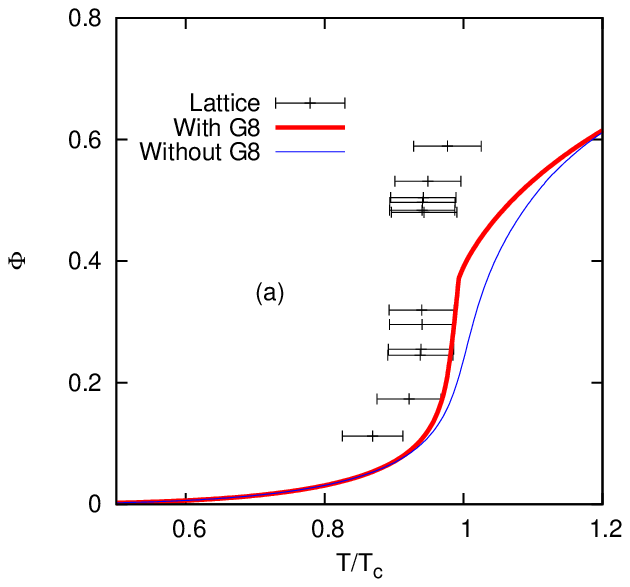}
 \includegraphics[width=0.5\textwidth]{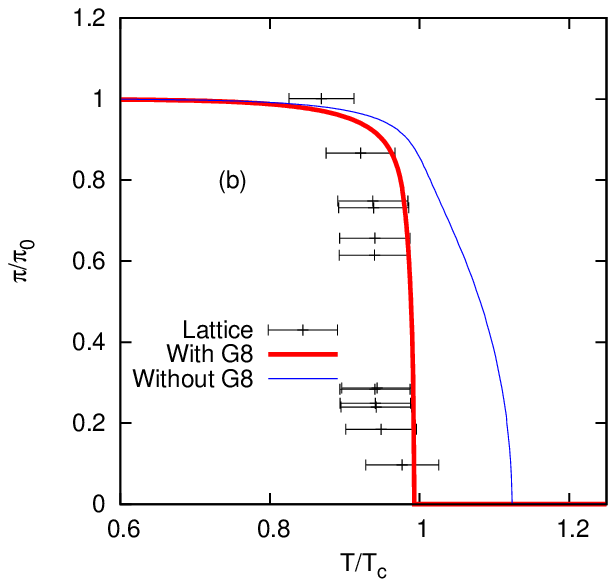}
\end{center}
\caption{(color online). $T$ dependence of 
(a) Polyakov-loop and (b) pion condensate 
at $\mu_{\rm I}=1.4\mu_{\rm c}=96$~[MeV] and $\mu_{\rm q}=0$. 
Here, $\pi$ is normalized by the value $\pi_0$ at zero $T$. 
See Fig.~\ref{T-dep-muq0} for the definition of lines and the LQCD data. 
  }
\label{T-dep-muq0-3}
\end{figure}

Next, we consider the pion-superfluidity phase by taking a case of 
$\mu_{\rm I}=1.4\mu_{\rm c}=96$~[MeV]. 
Figure~\ref{T-dep-muq0-3} presents $\Phi$ and $\pi$  
as a function of $T/T_\mathrm{c}$, 
where $\pi$ is normalized by the value $\pi_0$ at zero $T$. 
Again, the PNJL model with the scalar-type eight-quark interaction 
(the thick-solid curve) is consistent with the LQCD data compared with 
the PNJL model without the scalar-type eight-quark interaction 
(the thin-solid curve). The PNJL calculation on $\pi$ shows 
that the pion-superfluidity phase transition is of second order there.

\begin{figure}[htbp]
\begin{center}
 \includegraphics[width=0.45\textwidth]{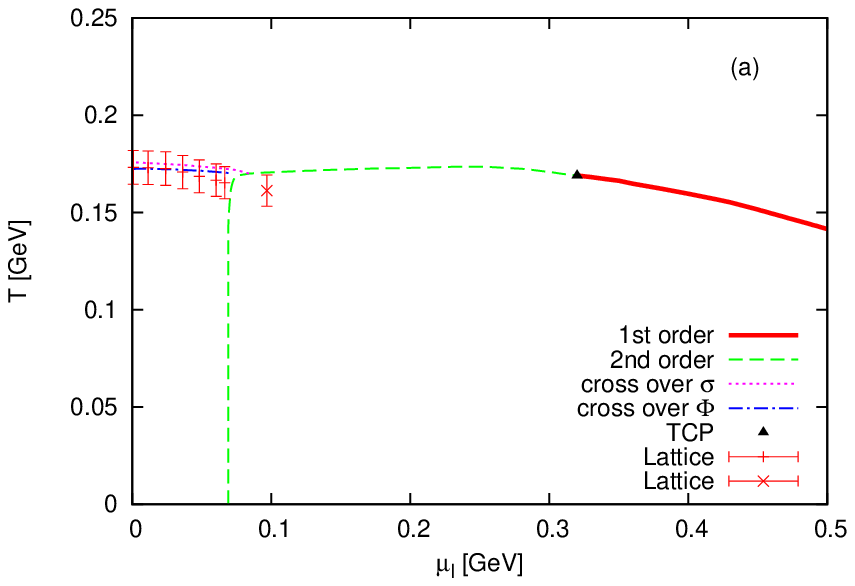}
 \includegraphics[width=0.45\textwidth]{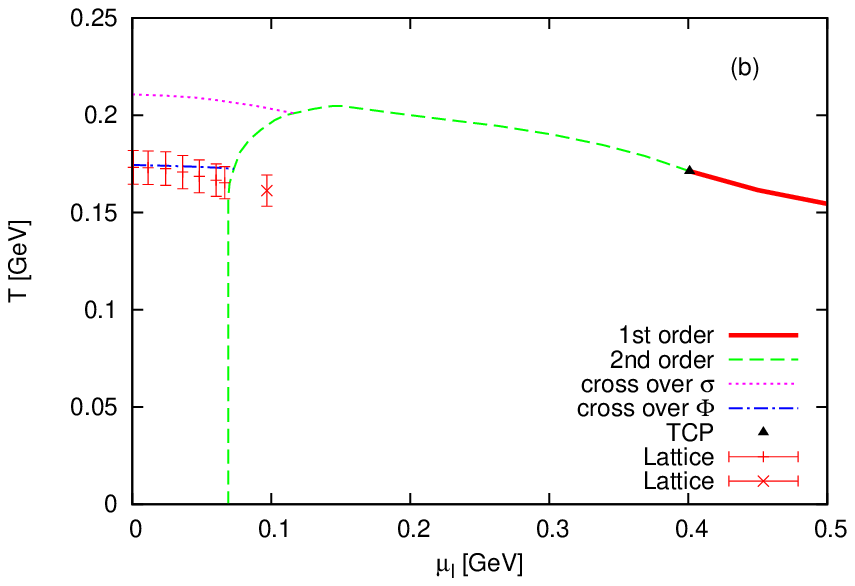}
\end{center}
\caption{(color online). Phase diagram 
in the $\mu_{\rm I}$-$T$ plane at $\mu_{\rm q}=0$ 
for the case (a) with and (b) without the scalar-type eight-quark interaction. 
The thick-solid (dashed) curve represents a first-order (second-order) 
pion-superfluidity phase transition. 
The dot-dashed (dotted) line means 
a deconfinement (chiral) crossover transition. 
At $\mu _{I}>M_{\pi}/2$, $\tilde{\chi} _{\Phi}$ has two peaks, so we do not 
plot any deconfinement crossover transition line there. 
Meanwhile, the chiral crossover transition line (dotted line) 
terminates at TCP. 
LQCD on the chiral and deconfinement crossover transitions 
are represented by a plus ($+$) 
symbol, while LQCD on the second-order 
pion-superfluidity transition and the deconfinement crossover transition 
are shown by 
a cross ($\times$) symbol. 
See Fig.~\ref{T-dep-muq0} for more information on LQCD data. 
}
\label{mu-I-T-q0}
\end{figure}

Thus, the PNJL model with the scalar-type eight-quark interaction is 
consistent with the LQCD data, indicating 
that the model is more reliable than the original PNJL model without 
the eight-quark interaction.  
Figure~\ref{mu-I-T-q0} shows the phase diagram in the $\mu_{\rm I}$-$T$ 
plane at $\mu_{\rm q}=0$. 
Panels (a) and (b) present results of the PNJL calculations with and without 
the eight-quark interaction, respectively. 
The thick-solid curve shows a first-order pion-superfluidity 
phase transition, 
while the dashed line indicates a second-order pion-superfluidity 
phase transition. 
A meeting point between the two lines is a tricritical point (TCP) by 
definition. 
The dot-dashed (dotted) line stands for a deconfinement (chiral) 
crossover transition. 
In panel (a), the two crossover transitions almost agree with each other. 
In LQCD, meanwhile, the agreement is perfect, as represented by a plus ($+$) 
symbol with 10 \% error bar. 
LQCD data on the pion-superfluidity transition is also shown by a cross 
($\times$) symbol with 10 \% error bar. 
Comparing the PNJL results with LQCD data, we can confirm that the PNJL model 
with the eight-quark interaction is more consistent with the LQCD data 
than that without the eight-quark interaction. 
The location of TCP is $(\mu_{\rm I},T)=(0.32~[{\rm GeV}],0.169~[{\rm GeV}])$ 
for the PNJL model with the eight-quark interaction 
and $(\mu_{\rm I},T)=(0.401~[{\rm GeV}],0.171~[{\rm GeV}])$ for 
the PNJL model without the eight-quark interaction. 
Thus, the eight-quark interaction is a sizable effect also on the location of 
TCP. 

\subsection{Phase structure in the $\mu_{\rm I}$-$\mu_{\rm q}$ plane 
at $T=0$}
\label{the IQ plane}

\begin{figure}[htbp]
\begin{center}
\includegraphics[width=0.45\textwidth]{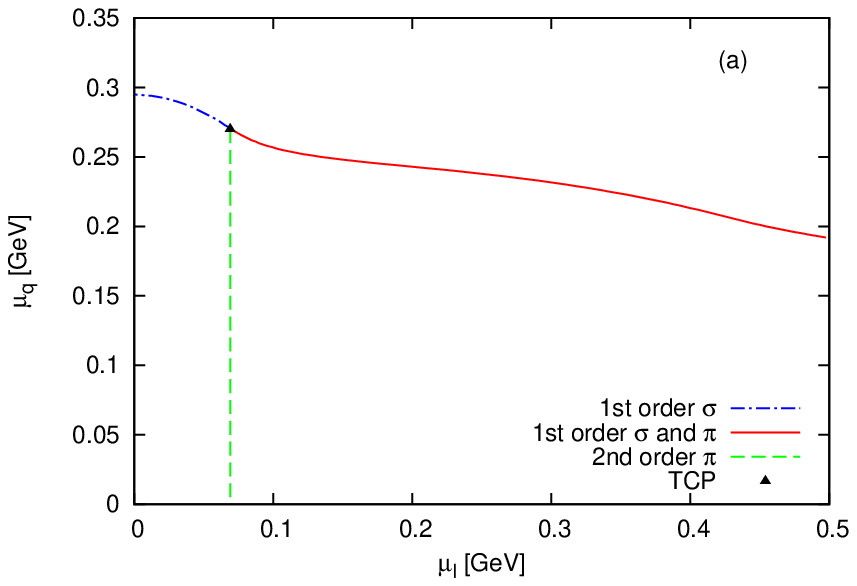}
\includegraphics[width=0.45\textwidth]{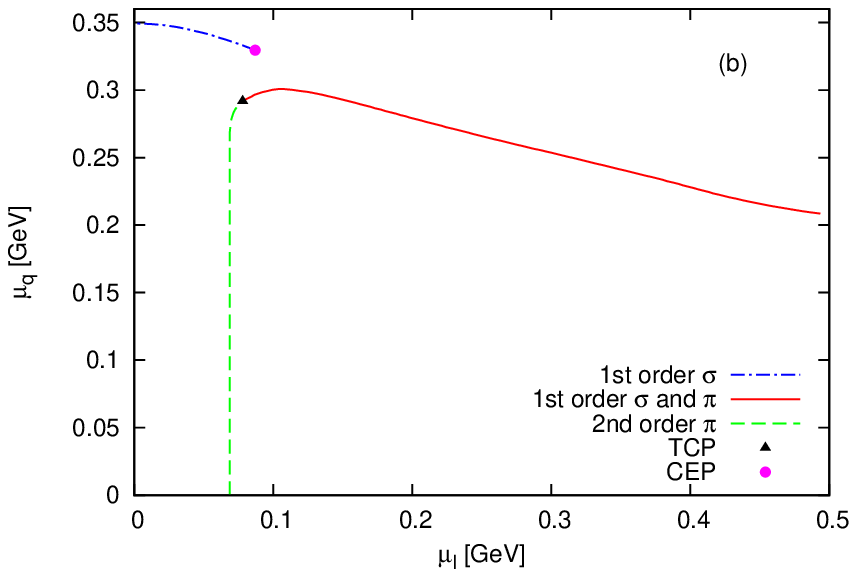}
\end{center}
\caption{(color online). Phase diagram 
in the $\mu_{\rm I}$-$\mu_{\rm q}$ plane at $T=0$ 
for the case (a) with and (b) without the eight-quark interaction. 
The solid line represents a coexistence line of first-order 
chiral and pion-superfluidity phase transitions, while 
the dot-dashed line shows a first-order chiral phase-transition line. 
The dashed line stands for 
a second-order pion-superfluidity phase transition.  
}
\label{Phase-diagram-IQ}
\end{figure}

In the $\mu_{\rm I}$-$\mu_{\rm q}$ plane at $T=0$, 
the thermodynamic potential of the PNJL model 
is reduced to that of the NJL model:
\bea
\Omega &=& 
-6\sum_{i=\pm}\int \frac{d^3{\rm p}}{(2\pi)^3}
         \Bigl[ E_{i}({\rm p}) - (E_i-\mu _{q})\theta (\mu_{q}-E_i) \Bigl]
         \nonumber\\
       &~& +G_{\rm s}[\sigma^2 +\pi^2]
       + 3G_{8}(\sigma ^{2} + \pi^{2})^{2}. 
\label{Omega-IQ0}
\eea
When $\mu_{\rm I} \le M_{\pi}/2$, $\pi=0$ and $M \approx 330$~MeV, so that 
$E_{\pm}= E \pm \mu_{\rm I} \ge M - {M_{\pi}\over{2}} \approx 260$~MeV. 
Hence, when $\mu_{\rm q} < 260$~MeV, $\Omega$ is reduced to 
\begin{equation}
\Omega = 
G_{\rm s}\sigma^{2}+3G_{8}\sigma ^{4}
-12 \int \frac{d^3{\rm p}}{(2\pi)^3} E({\rm p}). 
\label{}
\end{equation}
Therefore, $\Omega$ does not depend on $\mu_{q}$ and $\mu_{\rm I}$ for $\mu_{\rm I} < M_{\pi}/2$ and $\mu_{\rm q} < 260$~MeV, indicating that 
no phase transition occurs there. 
In other words, there is a possibility that 
a chiral phase transition takes place when $\mu_{\rm q} > 260$~MeV. 
This is realized, as shown later in Fig.~\ref{Phase-diagram-IQ}. 
For $\mu_{\rm I}=M_{\pi}/2$, more careful discussion is necessary, 
since it is a boundary of the normal phase in which $\pi=0$. 
This is discussed below.

In the normal-phase region at $\mu_{\rm q} < 260$~MeV, 
the curvature of $\Omega$ in the $\pi$-direction 
is obtained by \cite{He}  
\begin{eqnarray}
{\partial ^2\Omega\over{{\partial \pi}^2}}
=2G_\pi 
-48{G_{\pi}}^2\int {d^3{\bf p}\over{(2\pi)^3}}{E({\bf p})\over{E({\bf p})^2-\mu_{\rm I}^2}}
\equiv& f(\mu_{\rm I}),~~~~~~
\label{RPA1}
\end{eqnarray}
with 
\bea
G_{\pi}=-{1\over{2}}{\partial N\over{\partial \pi}} .
\eea
Thus, $f(\mu_{\rm I})$ does not depend on $\mu_{q}$. 
Here, an effect of the eight-quark interaction appears only through $M$ and $G_{\pi}$.

In vacuum ($T=\mu_{\rm q}=\mu_{\rm I}=0$), 
the RPA function with external momentum $(q_0\neq 0,{\bf q}=0)$ is~\cite{Kashiwa1,Fu,He}
\begin{eqnarray}
2G_\pi
-48{G_\pi}^2\int {d^3{\bf p}\over{(2\pi)^3}}{E({\bf p})\over{E({\bf p})^2-q_0^2/4}}
=f\left({q_0\over{2}}\right), 
\label{RPA2}
\end{eqnarray}
and the pion mass $M_\pi$ is determined by the condition $f(M_\pi/2)=0$. 
We then find for $\mu_{\rm I}=M_\pi/2$ that 
\begin{eqnarray}
{\partial ^2\Omega\over{{\partial \pi}^2}}
&=&f\left({M_\pi\over{2}}\right)=0 .
\label{RPA3}
\end{eqnarray}
Thus, the curvature of $\Omega$ at $\mu_{\rm I}=M_\pi/2$ is zero 
in the $\pi$ directions, indicating that 
a second-order pion-superfluidity phase transition takes place 
at $\mu_{\rm I}=M_{\pi}/2$ when $\mu_{\rm q} < 260$~MeV. 
This point will be confirmed later 
in the phase diagram of Fig.~\ref{Phase-diagram-IQ}(a) where 
the second-order pion-superfluidity phase transition line 
(dashed line) is a straight line. 

\begin{figure}[htbp]
\begin{center}
\includegraphics[width=0.45\textwidth]{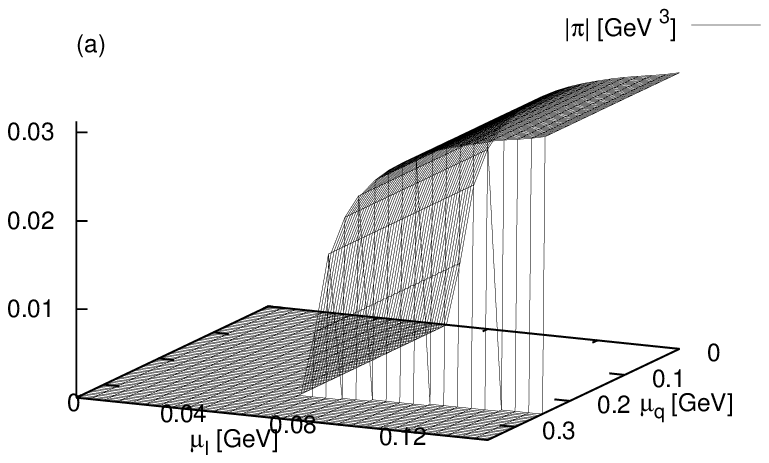}
\includegraphics[width=0.45\textwidth]{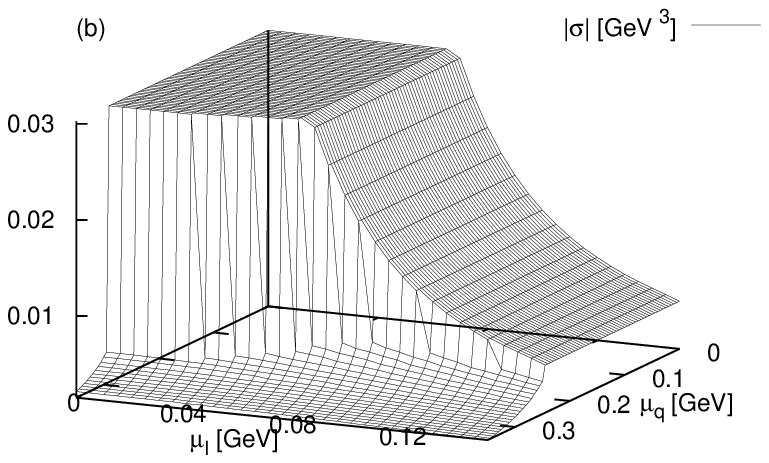}
\includegraphics[width=0.45\textwidth]{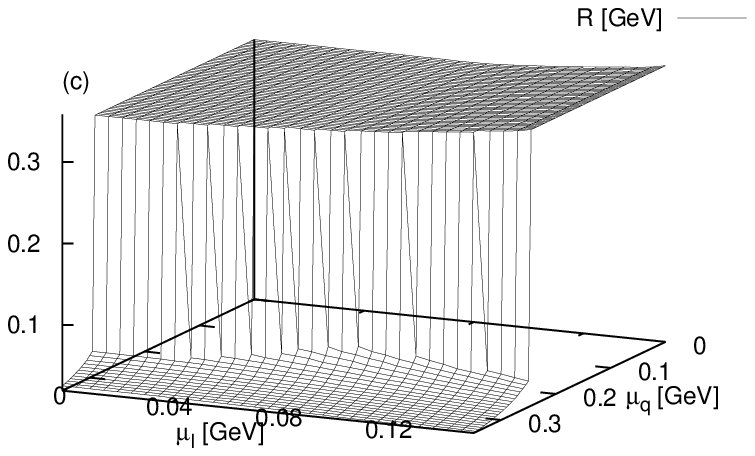}
\end{center}
\caption{(color online). Order parameters 
(a) $\pi$, (b) $\sigma$ and (c) $R$ 
as a function of $\mu_{\rm I}$ and $\mu_{\rm q}$. 
 }
\label{R-IQ}
\end{figure}

Next we consider both regions of $\mu_{\rm I} \le M_{\pi}/2$ and 
$\mu_{\rm I} > M_{\pi}/2$. 
Figure~\ref{R-IQ} shows 
$|\pi|$, $|\sigma|$ and $R=\sqrt{M^2+N^2}$ 
as a function of $\mu_{\rm I}$ and $\mu_{\rm q}$. 
The pion condensate $\pi$ is zero 
at $\mu_{\rm I}<M_{\pi}/2$, but nonzero 
at $\mu_{\rm I} > M_{\pi}/2$, as expected. Therefore, 
the former region is the normal ($I_3$-symmetric) phase 
and the latter region is 
the pion-superfluidity ($I_3$-symmetry broken) phase.  
The order parameter $|\sigma|$ of the chiral symmetry is almost constant in 
the normal phase but goes down in the pion-superfluidity phase. 
The parameter $R$ is almost constant over the two phases, 
when $\mu_{\rm q} < 200$~MeV. 
When $\mu_{\rm q}> 200$~MeV, $R$ has a discontinuity in the $\mu_{\rm q}$ 
direction. Thus, 
the $\mu_{\rm q}$ dependence of $R$ at finite $\mu_{\rm I}$ is similar to 
that at $\mu_{\rm I}=0$ over a wide range of $\mu_{\rm I}$.

In the limit of $m_0=0$ and $\mu_{\rm I}=0$, 
the chiral symmetry is an exact symmetry. In this situation, the thermodynamic 
potential of \eqref{eq:E12-pi} is a function of $R$, and $R$ is a function 
of $\sqrt{\sigma^2+\pi^2}$. This means that 
$R$ or $\sqrt{\sigma^2+\pi^2}$ is an order parameter of the chiral symmetry. 
When $m_0$ and/or $\mu_{\rm I}$ is finite, the chiral symmetry is not an exact 
symmetry anymore. However, the fact that 
the $\mu_{\rm q}$ dependence of $R$ at finite $\mu_{\rm I}$ is similar to 
that at $\mu_{\rm I}=0$ means that the chiral symmetry is preserved 
with good accuracy. 
This is understood as follows.

Over the normal and pion-superfluidity phases, we have 
\begin{eqnarray}
E_{\pm}
&=&\sqrt{(E\pm \mu_{\rm I})^{2}+N^{2}} \nonumber\\
&=&\sqrt{p^2 + R^2 \pm 2E\mu_{\rm I} + \mu_{\rm I} ^{2}} .
\end{eqnarray}
As shown in Fig.~\ref{R-IQ}, 
$R$ is about 330~MeV 
at $\mu_{\rm q} \la 200$~MeV and $\mu_{\rm I} < \Lambda = 631.5$~MeV. 
In the region, 
$E_{\pm}$ is well approximated by 
$\sqrt{p^2 + R^2 + \mu_{\rm I}^2}$, because 
$p^2 + R^2 + \mu_{\rm I}^2 \gg 2 E\mu_{\rm I}$. 
When $\mu_{\rm q} \ga 200$~MeV, $R$ is small and hence 
$E_{\pm}$ is approximated by 
$\sqrt{p^2 + \mu_{\rm I}^2}$. 
Therefore, 
$\Omega$ is a function of $R$ or $\sqrt{\sigma^2+\pi^2}$ 
with good accuracy; here, 
note that $R \approx \sqrt{\sigma^2+\pi^2}$ because of $m_0 \ll R$. 
Thus, when $T$ is small, 
the thermodynamics at finite $\mu_{\rm q}$ and $\mu_{\rm I}$ is 
controlled by an approximate order parameter $R$ of the chiral symmetry 
over both the normal ($\pi=0$) and the pion-superfluidity ($\pi \neq 0$) 
phase; the chiral symmetry is spontaneously broken when $R$ is finite, 
while it is restored when $R$ is zero. 
When $T \ga T_c$, $R$ is not large any more. Hence, $\sigma$ and $\pi$ 
work independently there, as shown later in subsection \ref{the space}.

Figure~\ref{Phase-diagram-IQ} presents the phase diagram in 
the $\mu_{\rm I}$-$\mu_{\rm q}$ plane at $T=0$. When $T=0$, 
the system is in the confinement phase because $\Phi=0$ there. 
So we consider the chiral and 
pion-superfluidity transitions only here. 
On the solid line, the first-order 
chiral and pion-superfluidity transitions coexist, 
while on the dot-dashed line 
only the first-order chiral transition takes place. 
The dashed line represents the second-order pion-superfluidity transition. 
In panel (a) where the eight-quark interaction is taken into account, 
the solid, dot-dashed and dashed lines meet at a point. This is a TCP, 
because the pion-superfluidity transition changes the order from 
first order to second order there, 
while the chiral transition keeps first order. 
Thus, there is no CEP in the $\mu_{\rm I}$-$\mu_{\rm q}$ plane at $T=0$. 
In panel (b) where the eight-quark interaction is switched off, 
the endpoint of the dot-dashed line is a CEP and 
a meeting point of the solid and dashed lines is a TCP by definition. 
Comparing the two panels, we can see that 
the eight-quark interaction changes the phase diagram qualitatively.

Recently, it was shown in Ref. ~\cite{Skokov} that 
the $\Lambda$ dependence of $\Omega$ may change the order of 
the phase transition in the mean field level. We then 
investigate the $\Lambda$ dependence of 
the phase diagram in the $\mu_{\rm I}$-$\mu_{\rm q}$ plane at $T=0$.
In this procedure, 
we consider the four-quark and eight-quark interactions only. 
The parameters of the PNJL model are determined for each $\Lambda$ so as to 
reproduce the pion decay constant $f_{\pi}=93.3$~MeV and the pion mass 
$M_{\pi}=138$~MeV at vacuum 
and $T_{\rm c}=173 \pm 8$~MeV 
at finite temperature with no $\mu_{\rm q}$ 
and $\mu_{\rm I}$~\cite{Karsch4,Karsch3,Kaczmarek2}; 
note that $T_{\rm c}$ is a much stronger constraint on 
$G_8$ than $M_{\sigma}$, 
since $M_{\sigma}$ has a large error bar~\cite{Sakai2}. 
This parameter fitting is exactly the same as 
that in Sec.~\ref{PNJL} to determine the parameter set A.

We vary $\Lambda$ from 573 to 651.5~MeV. 
The upper and the lower bound of $\Lambda$ are determined as follows. 
The QCD sum rule yields the lower and the upper bound of $|\sigma|$ 
as $|\sigma|=(225\pm 25{\rm MeV})^3$~\cite{Gasser,Reinders}. 
The absolute value of the chiral condensate, $|\sigma|$, 
increases as $\Lambda$ goes up, and 
reaches the upper bound of $|\sigma|$ when $\Lambda = 651.5$~MeV. 
Thus, $\Lambda = 651.5$~MeV is the upper bound of $\Lambda$. 
Meanwhile, the lower bound of $\Lambda$ is determined 
by not the lower bound of $|\sigma|$ but the fact that 
no parameter set can reproduce $f_{\pi}=93.3$~MeV and $M_{\pi}=138$~MeV
simultaneously when $\Lambda < 573$~MeV. 
Although this fact is found by numerical calculations, it 
can be understood with reasonable approximations. 
At zero temperature, the thermodynamic potential of the PNJL model is reduced to 
that of the NJL model, as shown in (\ref{Omega-IQ0}). 
In the NJL model, the pion mass is obtained by 
\bequ
M_{\pi}^2
=
\frac{-4m_0\sigma}{(M-m_0)MI(M,M_{\pi})} 
\label{mpi1}
\eequ
with 
\bequ
I(M,M_{\pi})
=
\frac{8N_fN_c}{2\pi^2}\int^{\Lambda}_{0}
\frac{p^2dp}{\sqrt{p^2+M^2}[4(p^2+M^2)-M_{\pi}^2]}, 
\label{mpi2}
\eequ
where $N_f$ and $N_c$ are the numbers of colors and flavors, respectively, and 
$N_f=2$ and $N_c=3$ in the present case. 
Since $M \gg m_0$ in \eqref{mpi1}, and 
$M^2 \gg M_{\pi}^2$ in \eqref{mpi2}, 
we then neglect $m_0$ in \eqref{mpi1} and $M_{\pi}$ in \eqref{mpi2} 
in order to understand the mathematical 
structure of \eqref{mpi1} and \eqref{mpi2}. 
Using the approximate equations and the Gell-Mann-Oakes-Renner relation, 
we have 
\begin{equation}
\frac{N_fN_c}{4\pi^2}x^2\left[
\ln \frac{1+\sqrt{1+x^2}}{x}-\frac{1}{\sqrt{1+x^2}}
\right] =\frac{f^2_{\pi}}{\Lambda^2} ,
\label{mpi3}
\end{equation}
where $x=M/\Lambda$. The left-hand side of \eqref{mpi3} has a maximum at 
$x=0.97$, while the right-hand side increases monotonously as 
$\Lambda$ goes down. This means that there exists a 
lower bound of $\Lambda$ that 
satisfies \eqref{mpi3}. The lower bound is $\Lambda=573$~MeV, 
although $\sigma$ obtained there is within the constraint 
$|\sigma|=(225\pm 25{\rm MeV})^3$ from the QCD sum rule.

\begin{figure}[htbp]
\begin{center}
\includegraphics[width=0.45\textwidth]{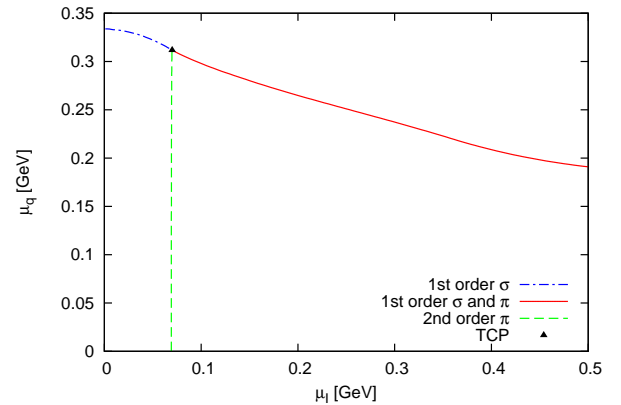}
\end{center}
\caption{
(color online). Phase diagram 
in the $\mu_{\rm I}$-$\mu_{\rm q}$ plane at $T=0$ 
calculated with set A'. 
See Fig.~\ref{Phase-diagram-IQ} for the definition of lines. 
}
\label{IQ-another-parameter}
\end{figure}

Table \ref{another-para} presents three parameter sets, A, A' and A", 
obtained by the above procedure. 
Set A is the original parameter set mentioned in Sec.~\ref{PNJL}, 
set A' is an example of the parameter sets near the lower bound of 
$\Lambda$, and set A" is the parameter set at the upper bound of $\Lambda$.
The value of $\Lambda$ in set A" is slightly larger than that in set A. 
This indicates that set A" yields qualitatively 
the same phase diagram as set A. Actually, we have confirmed this 
with numerical calculations. 
Meanwhile, the phase diagram calculated with set A' is shown 
in Fig.~\ref{IQ-another-parameter}. 
The phase structure shows no 
qualitative difference from the result of set A 
in Fig.~\ref{Phase-diagram-IQ}(a), although 
the first-order 
chiral transition line (dot-dashed line) 
and the pion-superfluidity phase-transition line (solid line)
are slightly shifted down 
by decreasing $\Lambda$. 
Furthermore, we have confirmed 
that the phase diagram does not change qualitatively 
near the lower bound. 
Thus, the order of the phase-transition is not changed by varying 
$\Lambda$ in the range $573 < \Lambda < 651.5$~MeV. 
As a property of the parameter sets near the lower bound of $\Lambda$, 
$G_8$ is quite large. This means that 
the higher-order multiquark interactions 
than the eight-quark interaction may not be negligible there. 
However, this sort of analyses is beyond the scope 
of the present work.

\begin{table}[h]
\begin{center}
\begin{tabular}{lcccc}
\hline \hline
Set~&$G_s$ & $G_8$ & $m_0$ & $\Lambda$
\\
\hline
A'&~$5.755~[{\rm GeV}^{-2}]$~& ~$1264.2~[{\rm GeV}^{-8}]$~ & ~$5.77~[{\rm MeV}]$~ & ~$580~[{\rm MeV}]$~ \\
A&~$4.673~[{\rm GeV}^{-2}]$~& ~$452.12~[{\rm GeV}^{-8}]$~ & ~$5.5~[{\rm MeV}]$~ & ~$631.5~[{\rm MeV}]$~ \\
A''&~$4.295~[{\rm GeV}^{-2}]$~& ~$351.32~[{\rm GeV}^{-8}]$~ & ~$5.31~[{\rm MeV}]$~ & ~$651.5~[{\rm MeV}]$~ \\
\hline \hline
\end{tabular}
\caption{
Cutoff dependence of parameters. 
Here, $T_0=203$~MeV for the set A', $T_0=212$~MeV for the set A and $T_0=217$~MeV for the set A''.
}
\label{another-para}
\end{center}
\end{table}

\subsection{Phase structure in the $\mu_{\rm q}$-$T$ 
plane at $\mu_{\rm I}=0$ }
\label{the QT plane}

\begin{figure}[htbp]
\begin{center}
 \includegraphics[width=0.4\textwidth]{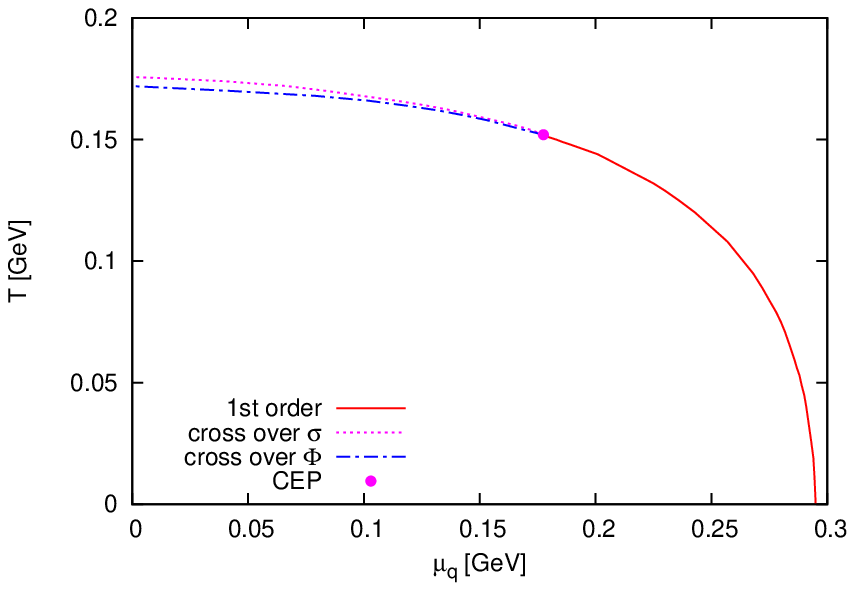}
\end{center}
\caption{(color online). Phase diagram 
in the $\mu_{\rm q}$-$T$ plane at $\mu_{\rm I}=0$. 
The solid line is a coexistence line of first-order 
chiral and deconfinement phase transitions. 
The dashed  line stands for the   chiral crossover transition, while  
the dot-dashed line does for the deconfinement  crossover transition. 
Here, the eight-quark interaction is taken into account in the PNJL model. 
}
\label{mu-q-T-I0}
\end{figure}

The phase diagram in the $\mu_{\rm q}$-$T$ plane at $\mu_{\rm I}=0$
is shown in Fig.~\ref{mu-q-T-I0}. 
The solid curve shows a coexistence line of 
first-order chiral and deconfinement phase transitions that ends  
at $(\mu_{\rm q},T)=(178~[{\rm MeV}], 152~[{\rm MeV}])$. 
This point is a CEP by definition and 
is known to be of second-order~\cite{AY,Fujii}. 
In general, once a first-order phase transition takes place for some 
order parameter, the discontinuity propagates to other order parameters 
unless the parameters are zero~\cite{Barducci,Kashiwa5}. 
The coexistence between the first-order chiral and deconfinement phase transitions shown in Fig.~\ref{mu-q-T-I0} is a typical case of the coexistence theorem. 
The dot-dashed (dotted) line stands 
for a crossover deconfinement (chiral) transition. 
The crossover chiral and deconfinement transitions almost coincide with each other and end at the CEP. 
Thus, a CEP exists in the present model. 
This CEP survives, even if the eight-quark interaction is switched off. 
In the case of no eight-quark interaction, the CEP in the $\mu_{\rm q}$-$T$ plane at $\mu_{\rm I}=0$ moves to a CEP in the 
$\mu_{\rm I}$-$\mu_{\rm q}$ plane at $T=0$ 
of Fig.~\ref{Phase-diagram-IQ}(b), 
as $\mu_{\rm I}$ increases from zero. 
This behavior of CEP is changed a lot by the eight-quark interaction, 
as shown later in Fig.~\ref{I-Q-T-8}.

\begin{figure}[htbp]
\begin{center}
 \includegraphics[width=0.4\textwidth]{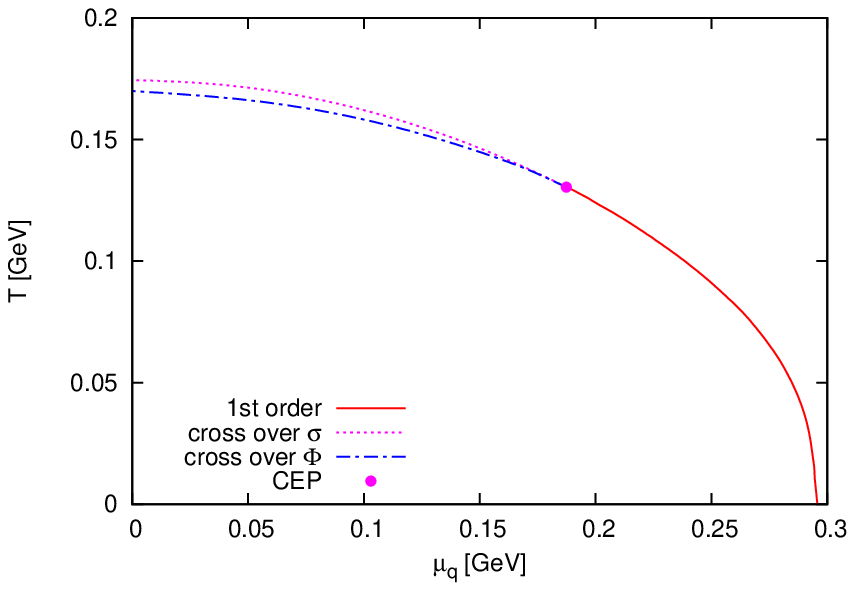}
\end{center}
\caption{
(color online). Effect of $\mu_{\rm q}$-dependent 
$T_{0}$ on the phase diagram 
in the $\mu_{\rm q}$-$T$ plane at $\mu_{\rm I}=0$.
In the PNJL calculation with the eight-quark interaction, 
parameter $T_{0}$ is replaced by 
$\mu_{\rm q}$-dependent parameter $T_0(\mu_{\rm q})$. 
See Fig.~\ref{mu-q-T-I0} for the definition of lines. 
}
\label{T0-running}
\end{figure}

In principle, the Polyakov-potential ${\cal U}$ 
depends on $\mu_{\rm q}$ as a consequence of 
the backreaction of the Fermion sector to the gluon sector. 
Particularly, the $\mu_{\rm q}$ dependence of the parameter $T_{0}$ 
in ${\cal U}$ is important and estimated by 
using renormalization group arguments~\cite{Schaefer}: 
\begin{eqnarray}
T_{0}(\mu_{\rm q})&=&T_{\tau}e^{-\frac{1}{\alpha_{0} b(\mu_{\rm q})}}
\end{eqnarray}
for $b(\mu_{\rm q})=29/(6\pi)-32\mu_{\rm q}^2/(\pi T_{\tau}^2)$ with 
$\alpha_{0}=0.304$ and $T_{\tau}=1.770{\rm [GeV]}$. 
Figure~\ref{T0-running} shows effects of $T_{0}(\mu_{\rm q})$ 
on the phase diagram 
in the $\mu_{\rm q}$-$T$ plane at $\mu_{\rm I}=0$. 
Comparing this figure with Fig.~\ref{mu-q-T-I0}, we can see that 
the effect dose not yield any qualitative change, but 
the location of CEP is moved from 
$(\mu_{\rm q},T)=(178~[{\rm MeV}], 152~[{\rm MeV}])$ to 
$(\mu_{\rm q},T)=(187~[\rm MeV],130~[\rm MeV])$. 
At small $T$, the effect becomes negligible, since ${\cal U}$ itself tends to 
zero as $T$ decreases.

\subsection{Phase structure in the $\mu_{\rm I}$-$\mu_{\rm q}$-$T$ space}
\label{the space}

\begin{figure}[htbp]
\begin{center}
 \includegraphics[width=0.5\textwidth]{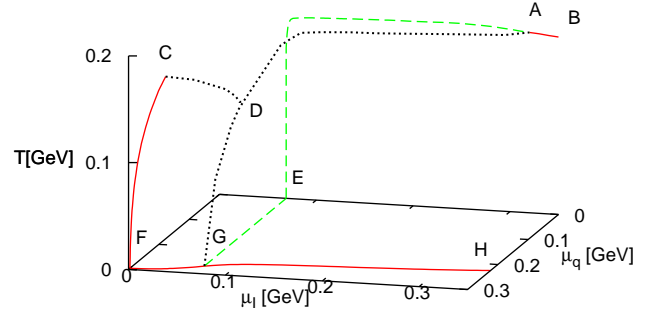}
\end{center}
\caption{(color online). Phase diagram 
in the $\mu_{\rm I}$-$\mu_{\rm q}$-$T$ space. 
Here, the eight-quark interaction is taken into account in the PNJL model. 
Properties of lines and areas 
are summarized in Table~\ref{Table-line}, while locations of 
points are summarized in Table~\ref{Table-point}.
  }
\label{I-Q-T-8}
\end{figure}

Figure~\ref{I-Q-T-8} presents the phase diagram 
in the $\mu_{\rm I}$-$\mu_{\rm q}$-$T$ space. 
In this space, TCP and CEP emerge not at points but on lines; 
precisely speaking, 
CEP appears on lines CD and DA, while TCP does on lines GD and DA. 
Thus, CEP moves from point C to A via D as $\mu_{\rm I}$ increases from zero. 
Meanwhile, TCP moves from point A to G via D 
as $\mu_{\rm q}$ increases from zero.

Line GE is a 
second-order pion-superfluidity transition line 
in the $\mu_{\rm I}$-$\mu_{\rm q}$ plane at $T=0$. 
A track of the line with respect to 
increasing $T$ becomes an area GEAD. 
Hence, the pion-superfluidity transition is 
second order on the area. 
Similarly, a track of line FG (GH) with respect to 
increasing $T$ becomes an area FGDC (GHBAD). 
In area FGDC, the chiral and deconfinement transitions are 
of first order, while the pion condensate is zero. 
In area GHBAD, 
all the chiral, deconfinement and pion-superfluidity transitions are 
of first order. 
The two areas smoothly connect to each other, indicating that 
the thermodynamics in these areas are controlled by $R$. 
Properties of lines and areas in Fig.~\ref{I-Q-T-8} 
are summarized in Table~\ref{Table-line}, while locations of 
points in Fig.~\ref{I-Q-T-8} are summarized in Table~\ref{Table-point}.

\begin{table}[h]
\begin{center}
\begin{tabular}{cccc}
\hline \hline
area&$\sigma $&$\pi $&$\Phi $\\
\hline
CDGF&~1st~&$\pi = 0$&1st\\
\hline
ABHGD&1st&1st&1st\\
\hline
ADGE&&2nd&\\
\hline \hline
\end{tabular}\\
\bigskip
\begin{tabular}{cccc}
\hline \hline
line&$\sigma $&$\pi $&$\Phi$\\
\hline
CF&1st&$\pi = 0$&1st\\
\hline
CD&CEP&$\pi =0$&CEP\\
\hline
FG&1st&$\pi = 0$&$\Phi = 0$\\
\hline
AD&CEP &TCP&CEP\\
\hline
DG&1st&TCP&1st\\
\hline
GE&&2nd&$\Phi = 0$\\
\hline
EA&&2nd&\\
\hline
GH&1st&1st&$\Phi = 0$\\
\hline \hline
\end{tabular}
\caption{
Properties of areas and lines in Fig.~\ref{I-Q-T-8}. 
The phrase ``1st" (``2nd") means 
that the phase transition either in the area or 
on the line is first (second) order. 
Blank means that no significant transition takes place there. 
}
\label{Table-line}
\end{center}
\end{table}

\begin{table}[h]
\begin{center}
\begin{tabular}{crcccccl}
\hline  \hline
point&(&$T [{\rm GeV}]$&,&$\mu _{\rm q} [{\rm GeV}]$&,&$\mu _{\rm I} [{\rm GeV}]$&)\\
\hline
A&(&0.169&,&0&,&0.320&)\\
\hline
B&(&0.166&,&0&,&0.350&)\\
\hline
C&(&0.152&,&0.178&,&0&)\\
\hline
D&(&0.136&,&0.190&,&0.084&)\\
\hline
E&(&0&,&0&,&0.069&)\\
\hline
F&(&0&,&0.295&,&0&)\\
\hline
G&(&0&,&0.270&,&0.069&)\\
\hline
H&(&0&,&0.223&,&0.350&)\\
\hline \hline
\end{tabular}
\caption{ Locations of points in Fig.~\ref{I-Q-T-8}. 
}
\label{Table-point}
\end{center}
\end{table}

\begin{figure}[htbp]
\begin{center}
 \includegraphics[width=0.5\textwidth]{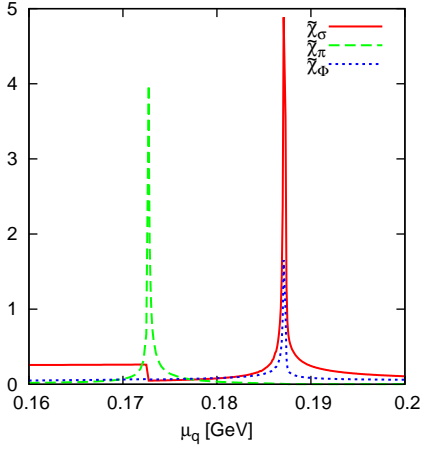}
\end{center}
\caption{(color online). 
Chiral, pion and Polyakov-loop susceptibilities 
as a function of $\mu_{\rm q}$ 
at $(\mu_{\rm I},T)=(0.075 [\textrm{GeV}],0.140 [\textrm{GeV}])$. 
Here, the eight-quark interaction is taken into account in the PNJL model. 
These are represented by the solid, dashed and dotted, respectively. 
See Ref.~\cite{Kashiwa1} for the definition of the susceptibilities. 
The $\tilde{\chi} _{\sigma}$ and $\tilde{\chi} _{\pi}$ are multiplied by $10^{-3}$ 
and $10^{-5}$, respectively, but $\tilde{\chi} _{\Phi}$ is not multiplied 
by any factor.   }
\label{near-D-1}
\end{figure}

Figure~\ref{near-D-1} presents the chiral 
susceptibility $\tilde{\chi}_{\sigma}$, the Polyakov-loop 
susceptibility $\tilde{\chi}_{\Phi}$ and the pion 
susceptibility $\tilde{\chi}_{\pi}$ as a function of $\mu_{\rm q}$ for the 
case of $(\mu_{\rm I},T)=(0.075~[\textrm{GeV}], 0.140~[\textrm{GeV}])$; 
these are plotted by 
the solid, dashed and dotted curves, respectively. 
The $\mu_{\rm q}$ dependence of these susceptibilities 
correspond to a line parallel to the $\mu_{\rm q}$ axis in Fig.~\ref{I-Q-T-8}. 
The susceptibilities $\tilde{\chi}_{\sigma}$ and $\tilde{\chi}_{\Phi}$ have 
peaks at the same position $\mu_{\rm q}=187$~MeV, indicating that 
the chiral and deconfinement transitions are second order there. 
This position corresponds to a point on line CD in Fig.~\ref{I-Q-T-8}. 
Meanwhile, $\tilde{\chi}_{\pi}$ has a peak at $\mu_{\rm q}=173$~MeV. This second-order 
critical point of the pion-superfluidity transition corresponds to a point 
on area ADGE in Fig.~\ref{I-Q-T-8}. 
As an interesting feature, 
$\tilde{\chi}_{\sigma}$ is discontinuous at $\mu_{\rm q}=173$~MeV. 
This property will be analyzed in Sec.~\ref{discontinuity}.

\begin{figure}[htbp]
\begin{center}
 \includegraphics[width=0.5\textwidth]{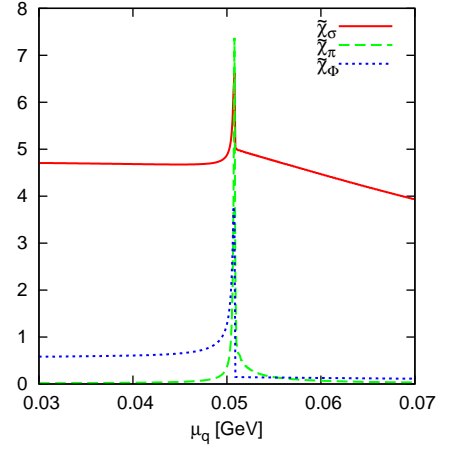}
\end{center}
\caption{(color online). 
Chiral, pion and Polyakov-loop susceptibilities 
as a function of $\mu_{\rm q}$ 
at $(\mu_{\rm I},T)=(0.100 [\textrm{GeV}],0.169 [\textrm{GeV}])$. 
Here, the eight-quark interaction is taken into account in the PNJL model. 
See Fig.~\ref{near-D-1} for the definition of lines. 
$\tilde{\chi} _{\sigma}$ and $\tilde{\chi} _{\pi}$ are multiplied by $1/20$ and 
$10^{-4}$, respectively, 
but $\tilde{\chi} _{\Phi}$ is not multiplied 
by any factor. 
  }
\label{near-D-2}
\end{figure}

Figure~\ref{near-D-2} shows $\tilde{\chi}_{\sigma}$, 
$\tilde{\chi}_{\Phi}$ and  $\tilde{\chi}_{\pi}$ as a function 
 of $\mu_{\rm q}$ for the case of 
$(\mu_{\rm I},T)=(0.100 [\textrm{GeV}],0.169 [\textrm{GeV}])$. 
All the susceptibilities have peaks at the same position 
$\mu_{\rm q}=51$~MeV, indicating that 
chiral, deconfinement and pion-superfluidity transitions of second order 
take place simultaneously there. This critical point 
corresponds to a point on line DA in Fig.~\ref{I-Q-T-8}. 
This is a TCP for $\pi$ and a CEP for $\sigma$ and $\Phi$. 
As an interesting feature, each of 
$\tilde{\chi}_{\sigma}$ and $\tilde{\chi}_{\Phi}$ has a kink at $\mu_{\rm q}=51$~MeV. 
This property will be analyzed in Sec.~\ref{discontinuity}.

Now, the phase diagram in the $\mu_{\rm I}$-$\mu_{\rm q}$-$T$ space is 
understood more precisely by considering the $\mu_{\rm q}$-$T$ plane 
at four values of $\mu_{\rm I}$: each belongs to 
any of four regions,  
(i) $\mu_{\rm I} < \mu_{\rm I}(G)=M_{\pi}/2$, 
(ii) $\mu_{\rm I}(G) < \mu_{\rm I} < \mu_{\rm I}(D)$, 
(iii) $\mu_{\rm I}(D) < \mu_{\rm I} < \mu_{\rm I}(A)$
and (iv) $\mu_{\rm I}(A) < \mu_{\rm I}$, 
where $\mu_{\rm I}(X)$ is a value of $\mu_{\rm I}$ at point X. 
The $\mu_{\rm q}$-$T$ phase diagram in region (i) 
is essentially equal to that in the $\mu_{\rm q}$-$T$ plane at 
$\mu_{\rm I}=0$, 
i.e., Fig.~\ref{mu-q-T-I0}, 
since $\pi$ is always zero there.

\begin{figure}[htbp]
\begin{center}
 \includegraphics[width=0.4\textwidth]{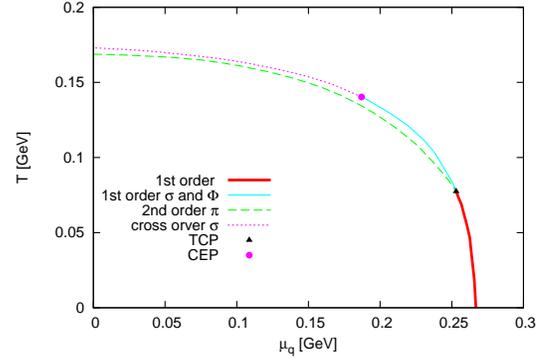}
\end{center}
\caption{(color online). Phase diagram 
in the $\mu_{\rm q}$-$T$ plane at $\mu_{\rm I}=75$~MeV. 
Here, the eight-quark interaction is taken into account in the PNJL model. 
}
\label{Q-T-2}
\end{figure}

The $\mu_{\rm q}$-$T$ phase diagram in region (ii) 
is a bit more complicated, as shown in Figure~\ref{Q-T-2} where 
$\mu_{I}=75$~MeV is taken as an example. 
In Fig.~\ref{Q-T-2}, 
the thick-solid line ending at TCP 
stands for a coexistence line of 
first-order chiral, deconfinement and pion-superfluidity transitions. 
This is a natural result of the coexistence theorem of the first-order 
phase transition~\cite{Barducci,Kashiwa5}. 
Meanwhile, on the thin-solid line between TCP and CEP, 
first-order chiral and deconfinement transitions coexist, but 
any first-order pion-superfluidity transition does not take place, 
because $\pi$ is zero above the dashed line starting from TCP that 
represents a second-order pion-superfluidity transition.

\begin{figure}[htbp]
\begin{center}
 \includegraphics[width=0.4\textwidth]{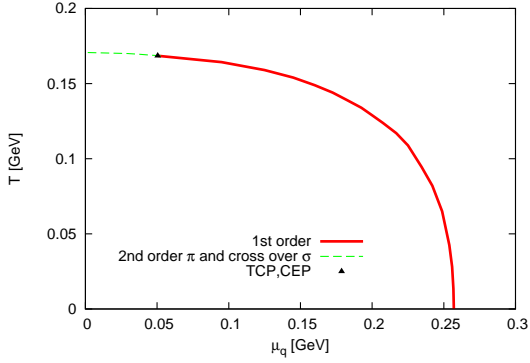}
\end{center}
\caption{(color online). Phase diagram 
in the $\mu_{\rm q}$-$T$ plane at $\mu_{\rm I}=100$~MeV. 
Here, the eight-quark interaction is taken into account in the PNJL model. 
The thick-solid line represents a coexistence line of 
the first-order chiral, deconfinement and 
pion-superfluidity transitions. 
On the dashed line, a second-order pion-superfluidity transition 
and a crossover chiral transition occur simultaneously. 
  }
\label{Q-T-3}
\end{figure}

The $\mu_{\rm q}$-$T$ phase diagram in region (iii) is 
simpler than that in region (ii). 
Fig.~\ref{Q-T-3} presents the $\mu_{\rm q}$-$T$ plane at 
$\mu_{I}=100$~MeV belonging to region (iii). 
As shown by the thick-solid line, 
all the first-order chiral, deconfinement and 
pion-superfluidity transitions occur simultaneously there. 
A second-order pion-superfluidity transition 
and a crossover chiral transition occur on the dashed line start 
from a point shown by triangle. This point is a TCP for $\pi$ and 
a CEP for $\sigma$ by definition. The point 
corresponds to a point on line DA in Fig.~\ref{I-Q-T-8}. 

The $\mu_{\rm q}$-$T$ phase diagram in region (iv) 
is simple and easily imaginable from Fig.~\ref{I-Q-T-8}. In this region, 
only a coexistence line of 
first-order chiral, deconfinement and pion-superfluidity transitions exists.

\subsection{Properties of susceptibilities}
\label{discontinuity}
Properties of the susceptibilities near 
the second-order pion-superfluidity transition line, CEP and TCP 
are investigated. 

For simplicity, we take the following shorthand notation 
for the curvature matrix $C$ of \eqref{curvature}: 
\begin{equation}
C=
\left(
\begin{array}{cc}
c_{\pi\pi}&A\\
A^{\rm T}&K
\end{array}
\right) ,
\label{C-short}
\end{equation}
where $A=(c_{\pi\sigma},c_{\pi\Phi},c_{\pi {\bar \Phi}})$, 
$A^{\rm T}$ is the transverse of $A$, and the matrix $K$ is expressed by
\begin{equation}
K=
\left(
\begin{array}{ccc}
c_{\sigma\sigma}&c_{\sigma\Phi}&c_{\sigma\bar{\Phi}}\\
c_{\Phi\sigma}&c_{\Phi\Phi}&c_{\Phi\bar{\Phi}}\\
c_{\bar{\Phi}\sigma}&c_{\bar{\Phi}\Phi}&c_{\sigma\bar{\Phi}}\\
\end{array}
\right) .
\end{equation}
As shown in \eqref{eq:E12-pi}, $\Omega$ is an even function of $\pi$. 
Noting that $c_{XY}$ ($X, Y=\pi,\sigma, \Phi, \bar{\Phi}$) are 
proportional to $\partial^2 \Omega/ \partial{X}\partial{Y}$; therefore, 
we can find that $c_{\pi\pi}$ 
and $c_{xy}$ for $x, y=\sigma, \Phi, \bar{\Phi}$ are $\pi$-even, while 
$c_{x\pi}$ and $c_{\pi y}$ for $x, y=\sigma, \Phi, \bar{\Phi}$ are $\pi$-odd. 

\begin{figure}[htbp]
\begin{center}
 \includegraphics[width=0.4\textwidth]{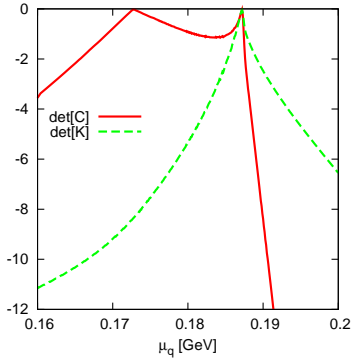}
\end{center}
\caption{(color online). 
$\mu_{\rm q}$ dependence of ${\rm det}[C]$ and ${\rm det}[K]$
at $T=0.14~{\rm [GeV]}$ and $\mu_{\rm I}=0.075~{\rm [GeV]}$. 
The solid (dashed) line stands for ${\rm det}[C]$ (${\rm det}[K]$).
Here, the eight-quark interaction is taken into account in the PNJL model.
The ${\rm det}[C]$ is multiplied by $6\times 10^{2}$.
}
\label{det-1}
\end{figure}

First, we consider the normal ($\pi=0$) phase 
including the second-order pion-superfluidity 
transition line. Since $\pi=0$ in this phase, 
the $\pi$-odd quantities $c_{x\pi}$ and $c_{\pi y}$ for $x, y=\sigma, \Phi, \bar{\Phi}$ are zero.  
Therefore, $C$ is reduced to 
\begin{equation}
C =
\left(
\begin{array}{cc}
c_{\pi\pi}&0\\
0&K
\end{array}
\right) .
\label{C-normal}
\end{equation}
Equation \eqref{C-normal} shows the following properties. 
\begin{enumerate}
	\item 
	On the second-order pion-superfluidity transition line, 
the curvature $c_{\pi\pi}$ in the $\pi$ direction 
is zero by definition of the second-order transition. 
Therefore, ${\rm det}[C]=0$. 
This indicates that $\tilde{\chi}_{\pi}={\rm det}[K]/{\rm det}[C]$ 
diverges on the transition line, since ${\rm det}[K]$ is not zero in general. 
    \item
If a CEP of the chiral phase transition appears in the normal phase, 
the determinant ${\rm det}[K]$ is zero at the CEP; see Ref.\cite{Fujii} 
for the details of this proof. Hence, $\tilde{\chi}_{\sigma}$ is divergent 
at the CEP because of ${\rm det}[C]=c_{\pi\pi}{\rm det}[K]=0$. 
\end{enumerate}
Properties 1 and 2 are understood clearly with 
numerical results shown in Fig.~\ref{near-D-1}. 
The peak of $\tilde{\chi}_{\pi}$ at $\mu_{\rm q}=\mu_{\rm q}^{\pi} = 173$~MeV
shows a second-order pion-superfluidity transition, 
while the peak of $\tilde{\chi}_{\sigma}$ 
at $\mu_{\rm q}=\mu_{\rm q}^{\sigma}=187$~MeV 
does a CEP in the normal  phase. 
Hence, the thermal system is in the normal phase ($\pi=0$) 
for $\mu_{\rm q}>\mu_{\rm q}^{\pi}$ and in the broken phase 
($\pi \neq 0$) for $\mu_{\rm q} < \mu_{\rm q}^{\pi}$. 
Figure~\ref{det-1} shows ${\rm det}[C]$ and ${\rm det}[K]$ as a function of 
$\mu_{\rm q}$ 
at $(\mu_{\rm I},T)=(0.075 [\textrm{GeV}],0.140 [\textrm{GeV}])$. 
It is found from this figure that 
${\rm det}[C]=0$ and ${\rm det}[K] \neq 0$ 
at $\mu_{\rm q}=\mu_{\rm q}^{\pi}$, while ${\rm det}[C]={\rm det}[K] =0$ 
at $\mu_{\rm q}=\mu_{\rm q}^{\sigma}$. Thus, properties 1 and 2 are confirmed 
to be true by the numerical results.

Next, we consider the broken ($\pi \neq 0$) phase. 
Fig.~\ref{near-D-1} is a good example. 
At $\mu_{\rm q}$ slightly smaller than $\mu_{\rm q}^{\pi}$, $\pi$ 
is small, because $\pi=0$ at $\mu_{\rm q}=\mu_{\rm q}^{\pi}$. 
Hence, any quantity can be expanded into a power series of $\pi$. 
After the expansion, the $\pi$-even quantities 
$c_{xy}$ ($x,y=\sigma,\Phi,{\bar \Phi}$) 
are of order $(\pi)^0$, 
while the $\pi$-odd quantities $c_{\pi y}$ and $c_{x\pi}$ 
($x,y=\sigma,\Phi,{\bar \Phi}$) are 
of order $(\pi)^1$. 
The entry $c_{\pi\pi}$ is of order $(\pi)^2$, 
as shown below.
The stationary condition \eqref{eq:SC} for $\pi$ is rewritten into 
\bea
0=\frac{\partial \Omega}{\partial \pi}
=\frac{\partial \Omega}{\partial \pi^2} \frac{d\pi^2}{d\pi}
=\frac{\partial \Omega}{\partial \pi^2} 2\pi,  
\eea
and hence   
\bea
\frac{\partial \Omega}{\partial \pi^2}=0 
\label{condition}
\eea
because of $\pi \neq 0$. Expanding the $\pi$-even function $\Omega$ 
into a power series of $\pi^2$, 
\bea
\Omega=\sum_n a_n \pi^{2n}, 
\eea
one can see from \eqref{condition} that 
\bea
a_1=0 .
\eea
Hence, 
$c_{\pi\pi} = T^2 \partial^2 \Omega/\partial \pi\partial \pi$ 
is of order $(\pi)^2$. 
Therefore, the matrix $C$ is the following property in the broken phase: 
\begin{enumerate}
	\item[3.] 
	At $\mu_{\rm q}$ slightly smaller than $\mu_{\rm q}^{\pi}$, 
	$c_{\pi\pi}$ is of order $(\pi)^2$, 
$A$ and $A^{\rm T}$ are of order $(\pi)^1$, and $K$ is of order 
$(\pi)^0$. 
\end{enumerate}

Now, we consider the reason why $\tilde{\chi}_{\sigma}$ is discontinuous  
at $\mu_{\rm q}=\mu_{\rm q}^{\pi}$ in Fig.~\ref{near-D-1}. 
The susceptibility $\tilde{\chi}_{\sigma}$ is expressed by 
\bea
\tilde{\chi}_{\sigma}=\frac{\Delta_{\sigma \sigma}}{{\rm det}[C]}, 
\label{chi-sigma}
\eea
where $\Delta_{\sigma \sigma}$ is the cofactor 
of entry $c_{\sigma \sigma}$ in the matrix $C$. 
Property 3 indicates that 
both $\Delta_{\sigma \sigma}$ and ${\rm det}[C]$ are 
of order $(\pi)^2$ in the broken phase at 
$\mu_{\rm q} < \mu_{\rm q}^{\pi}$, so that the left-hand limit of 
$\tilde{\chi}_{\sigma}$ as $\mu_{\rm q}$ approaches $\mu_{\rm q}^{\pi}$ 
is finite. 
As an important point, the $\pi$-odd quantities contribute 
to this left-hand limit. 
Meanwhile, the $\pi$-odd quantities are zero in the normal-phase at 
$\mu_{\rm q} > \mu_{\rm q}^{\pi}$, so that they do not contribute to 
the right-hand limit of $\Delta_{\sigma \sigma}$ and ${\rm det}[C]$ 
as $\mu_{\rm q}$ approaches $\mu_{\rm q}^{\pi}$. 
Thus, the right-hand limit of $\tilde{\chi}_{\sigma}$ is different from the 
left-hand limit of $\tilde{\chi}_{\sigma}$.

In Fig.~\ref{near-D-2}, all the susceptibilities, $\tilde{\chi}_{\sigma}$, 
$\tilde{\chi}_{\Phi}$ and  $\tilde{\chi}_{\pi}$, have peaks at the same position 
$\mu_{\rm q}=\mu_{\rm q}^{\pi}=51$~MeV. 
The divergence of $\tilde{\chi}_{\sigma}$ means that in \eqref{chi-sigma}, 
the denominator ${\rm det}[C]$ tends to zero faster than 
the numerator $\Delta_{\sigma \sigma}$ 
as $\mu_{\rm q}$ approaches $\mu_{\rm q}^{\pi}$ from the left-hand side. 
There is no guarantee that such a strong damping 
of ${\rm det}[C]$ also happens in the right-hand limit, because 
$\pi$-odd quantities $c_{\sigma y}$ and $c_{x\pi}$ are zero there. 
Actually, such a fast damping in the right-hand limit does not occur here, 
as shown by the numerical calculation. 
As $\mu_{\rm q}$ approaches $\mu_{\rm q}^{\pi}$, therefore, 
$\tilde{\chi}_{\sigma}$ is divergent in the left-hand limit, 
but finite in the right-hand limit. 
This is the reason why $\tilde{\chi}_{\sigma}$ has a kink 
at $\mu_{\rm q}=51$~MeV.

\begin{figure}[htbp]
\begin{center}
 \includegraphics[width=0.45\textwidth]{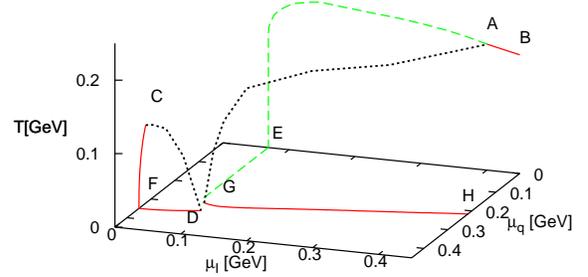}
\end{center}
\caption{ 
(color online). Phase diagram 
in the $\mu_{\rm I}$-$\mu_{\rm q}$-$T$ space calculated with 
the PNJL model with the four-quark interaction only. 
See Fig.~\ref{I-Q-T-8} for the definition of lines and areas, 
except line CD represents CEP and line AG stands for TCP.  
}
\label{souzu-4-point}
\end{figure}

The fast damping of ${\rm det}[C]$ 
in both the left- and the right-hand limit
happens only on point D in Fig.~\ref{I-Q-T-8}, as shown below.
Figure~\ref{on-D} presents $\tilde{\chi}_{\sigma}$, $\tilde{\chi}_{\Phi}$ and  $\tilde{\chi}_{\pi}$ 
as a function of $\mu_{\rm q}$ at $\mu_{\rm I}=0.08425$~GeV and $T=0.136$~GeV. 
All the susceptibilities diverge at $\mu_{\rm q}=0.190$~GeV. 
This peak corresponds 
to point D in Fig.~\ref{I-Q-T-8}. In this case, obviously, 
the susceptibilities have no kink. Therefore, 
${\rm det}[C]$ tends to zero faster than 
$\Delta_{\sigma \sigma}$ 
in both the right- and the left-hand limit. 
Point D is a meeting point of CEP and TCP. 
There is no guarantee that such a special critical point always happens. 
Actually, such a point does not appear if the eight-quark interaction is 
switched off, as shown in Fig.~\ref{souzu-4-point}; 
here, line CD (AG) represents  CEP (TCP) 
of the chiral (pion-superfluidity) phase transition and there is no 
meeting point between CEP and TCP.

\begin{figure}[htbp]
\begin{center}
 \includegraphics[width=0.5\textwidth]{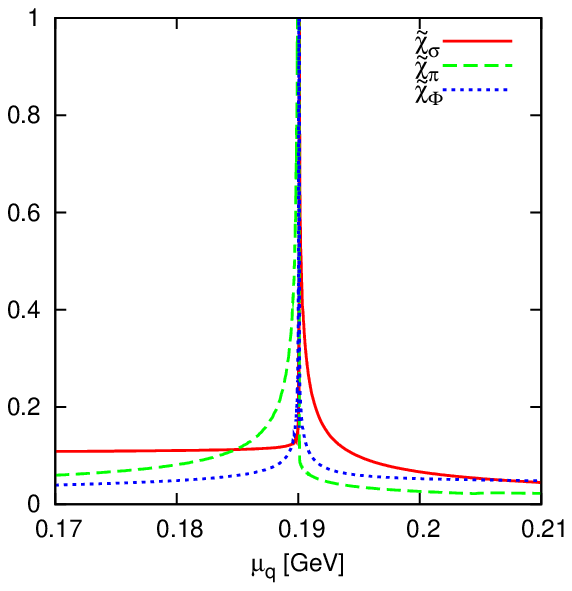}
\end{center}
\caption{(color online). 
Chiral, pion and Polyakov-loop susceptibilities 
as a function of $\mu_{\rm q}$ 
at $(\mu_{\rm I},T)=(0.08425 [\textrm{GeV}],0.136 [\textrm{GeV}])$. 
Here, the eight-quark interaction is taken into account in the PNJL model. 
See Fig.~\ref{near-D-1} for the definition of lines. 
The $\tilde{\chi} _{\sigma}$ and $\tilde{\chi} _{\pi}$ are multiplied by $5\times 10^{-4}$ 
and $2\times 10^{-4}$, respectively, but $\tilde{\chi} _{\Phi}$ is not multiplied 
by any factor. 
}
\label{on-D}
\end{figure}

\section{Summary}
\label{Summary}

Critical points such as CEP and TCP are important as 
indicators of the chiral, deconfinement and pion-superfluidity 
phase transitions in measurements at 
GSI, SPS, RHIC and LHC. 
In the measurements, $\mu_{\rm I}$ is 
not zero generally. We have then predicted the phase diagram of two-flavor QCD 
in the $\mu_{\rm I}$-$\mu_{\rm q}$-$T$ space 
by using the PNJL model with the scalar-type eight-quark interaction. 
The PNJL model with the scalar-type eight-quark interaction is 
consistent with the LQCD data~\cite{Kogut2} 
in the $\mu_{\rm I}$-$T$ plane at $\mu_{\rm q}=0$, 
while the original PNJL model without the scalar-type eight-quark interaction 
is not.

In the $\mu_{\rm q}$-$\mu_{\rm I}$-$T$ space, as shown in Fig.~\ref{I-Q-T-8}, 
a  CEP in the $\mu_{\rm q}$-$T$ plane at $\mu_{\rm I}=0$ moves to a TCP 
in the $\mu_{\rm I}$-$T$ plane 
$\mu_{\rm q}=0$ as $\mu_{\rm I}$ increases. 
Meanwhile, the TCP in the $\mu_{\rm I}$-$T$ plane at $\mu_{\rm q}=0$ 
moves to 
a TCP in the $\mu_{\rm q}$-$\mu_{\rm I}$ plane at $T=0$. 
When $\mu_{\rm I}<M_{\pi}/2$, the pion condensate $\pi$ is zero and hence 
a CEP exists but any TCP does not.  
When $M_{\pi}/2< \mu_{\rm I} \la 80$~MeV, a CEP and a TCP exist separately. 
And when $\mu_{\rm I} \ga 80$~MeV, they coexist.  
If the eight-quark interaction is switched off, 
a  CEP in the $\mu_{\rm q}$-$T$ plane at $\mu_{\rm I}=0$ moves to a CEP  
in the $\mu_{\rm q}$-$\mu_{\rm I}$ plane at $T=0$ as $\mu_{\rm I}$ increases; 
see Fig.~\ref{souzu-4-point}. 
Thus, the eight-quark interaction changes the QCD diagram 
qualitatively in the $\mu_{\rm q}$-$\mu_{\rm I}$-$T$ space.

When $T$ is small, the thermodynamics 
at finite $\mu_{\rm I}$ and $\mu_{\rm q}$ is controlled by 
$\sqrt{\sigma^2+\pi^2}$. 
The quantity $\sqrt{\sigma^2+\pi^2}$ 
is an approximate order parameter of the chiral symmetry over 
the $I_3$-symmetric ($\pi=0$) and $I_3$-symmetry broken ($\pi \neq 0$) 
phases.

\noindent
\begin{acknowledgments}
Authors thank P. de. Forcrand, A. Nakamura, K. Nagata and K. Kashiwa for useful discussions. 
H. K. also thanks M. Imachi, H. Yoneyama H. Aoki and M. Tachibana 
for useful discussions. 
Y. S acknowledges support by JSPS.
\end{acknowledgments}


\end{document}